\documentclass[preprint]{siamltexmm}

\usepackage{amsmath,amssymb}

\usepackage{hyperref, cite}

\title{\uppercase{Multi-Parameter Models of Innovation\\Diffusion on Complex Networks}\thanks{This work was funded under the EPSRC Energy Challenges for Complexity Science panel, grant EP/G059780/1.}}
\author{N.~J.~McCullen\footnotemark[2]
\and A.~M.~Rucklidge\footnotemark[2]
\and C.~S.~E.~Bale\footnotemark[3]
\and T.~J.~Foxon\footnotemark[4]
\and W.~F.~Gale\footnotemark[3]}

\begin{document}

\renewcommand{\thefootnote}{\fnsymbol{footnote}}
\footnotetext[2]{Department of Applied Mathematics, University of Leeds.}
\footnotetext[3]{Energy Research Institute, University of Leeds.}
\footnotetext[4]{School of Earth and Environment, University of Leeds.}
\renewcommand{\thefootnote}{\arabic{footnote}}

\maketitle

\newcommand{\slugmaster}{%
}

\pagestyle{myheadings}
\thispagestyle{plain}
\markboth{\uppercase{N.~J.~McCullen, A.~M.~Rucklidge, C.~S.~E.~Bale, T.~J.~Foxon and W.~F.~Gale}}{\uppercase{Models of Innovation Diffusion on Networks}}

\begin{abstract}
A model, applicable to a range of innovation diffusion applications with a strong peer to peer component, is developed and studied, along with methods for its investigation and analysis.
A particular application is to individual households deciding whether to install an energy efficiency measure in their home.
The model represents these individuals as nodes on a network, each with a variable representing their current state of adoption of the innovation.
The motivation to adopt is composed of three terms, representing personal preference, an average of each individual's network neighbours' states and a system average, which is a measure of the current social trend.
The adoption state of a node changes if a weighted linear combination of these factors exceeds some threshold.   
Numerical simulations have been carried out, computing the average uptake after a sufficient number of time-steps over many realisations at a range of model parameter values, on various network topologies, including random (Erd\H{o}s--R\'enyi), small world (Watts--Strogatz) and (Newman's) highly clustered, community-based networks. 
An analytical and probabilistic approach has been developed to account for the observed behaviour, which  explains the results of the numerical calculations.
\end{abstract}



\section{Introduction}

Social phenomena, such as the spread of a technological or behavioural innovation through communities, can be modelled as dynamical processes on networks \cite{granovetter1983threshold, alkemade2005strategies, delre2010will, castellano2009statistical, choi2010role, bikhchandani1992theory}.
Our model, introduced in \S \ref{sec:model}, builds on previous threshold diffusion models (e.g.\  \cite{watts2002simple, lee2006reconsideration, choi2010role}) by incorporating sociologically realistic factors, yet remains simple enough for mathematical insights to be developed.

An example of a particular application of this model is to the adoption of innovations related to energy behaviours and technologies by individual households.
These innovations are often not directly visible to an adopter's peers but communication of the benefits of adoption may occur through interaction between individuals.	
The decision to adopt is therefore based on multiple factors, taking into account not only individual preferences, but also whether or not an individual's social circle has adopted the innovation.
As such, the spread of the innovation will be influenced by the network of social contacts between individuals, including both social peers and wider social trends.
The real topologies of interpersonal networks are not known exactly and are constantly changing, further adding to the uncertainty in models of social phenomena. 
Given these challenges, useful models and methods are required that can assess the likely outcomes of particular scenarios, such as the effect of an intervention to persuade population to adopt an innovation, without being excessively sensitive to particular choices of parameter values of the model or the precise structure of the network.
Here, we present numerical investigations of the model, studying its  behaviour statistically, as well as presenting mathematical analysis that gives a deeper understanding of the observed uptake of the innovation over the network using probabilistic arguments.

The numerical methods used in \S \ref{sec:numerics} take the ensemble averages of simulation outcomes over many realisations, each with  different conditions, such as the initial seed of adopters or the precise details of the network structure. 
This is carried out over a range of parameter values and on different types of network topology in order to study both the stability of the model with respect to minor variations and the sensitivity to structural details of the models used.
These methods can, therefore, be applied to various dynamical models on a variety of networks to assess likely differences in outcomes for alternative scenarios in a statistical sense, rather than relying on individual predictions.

We take the applied dynamical systems approach of looking for lines in parameter space where the behaviour of the dynamical process changes.
Mathematical analysis is presented in \S \ref{sec:anal} explaining observations for the simulation results in certain simplified cases.
Specifically, assuming that all individuals are homogeneous in their model parameters, we can use probabilistic arguments for the conditions of the neighbourhoods of the individuals required to trigger uptake, and calculate the likelihood of success given a choice of parameters on certain types of networks.

We have found that the likelihood of success, defined as adoption by the majority of individuals, depends strongly on both the choice of model parameters and the topology of the network. 
Two particularly important factors are found to be the \emph{node degree}, i.e., the number of connections of the nodes with their neighbours, and the network's \emph{transitivity}, which is the degree of clustering related to the correlation between the neighbourhoods of connected individuals.
This makes it possible to assess what changes to the model would result in an increased probability of success, and how this can be interpreted in terms of policy decisions intent on maximising adoption in the real world.

\section{Modelling social behaviour as dynamical systems on networks}\label{sec:model}

A system of $N$ individuals (or groups such as households) can be represented by $N$ \emph{nodes} on a \emph{network}.
Possible interactions linking individuals are represented by the \emph{edges} on the network.
The network topology can be represented in the \emph{adjacency matrix} $A$, an $N^2$ matrix with entries:
\begin{equation}
A_{ij} = \left\{
  \begin{array}{l l}
    1 & \quad \text{if node $j$ influences $i$,}\\
    0 & \quad \text{otherwise.}\\
  \end{array} \right.
\end{equation}

Information or influence is passed along these edges, which could also be given weights to show the relative probability or strength of the interaction.
In this work all influences are taken to be of equal weight and symmetric in both directions.
States are assigned to the nodes, describing the properties of the individuals, and deterministic or probabilistic equations or rules can be used to describe the evolution of these states over time.

\subsection{A multi-parameter model for innovation uptake}

The following model describes the purchase or adoption of a single innovation (not considering competition between similar items) but could apply equally well to the spread of any property over a network, such as a rumour or behaviour. 
The model used in this work assigns each node $i$ with a binary variable representing their current state, $x_i=1,0$, indicating whether or not the individual has adopted the innovation.
The nodes in state $x=1$ are referred to as \emph{activated} nodes.
The initial state is a fixed proportion of activated nodes, randomly distributed across the network.
The decision to invest in the innovation is determined by the perceived usefulness, or \emph{utility}, to the individual (which can include subjective judgements).
When this utility outweighs the barriers to adoption (including financial costs), the individual $i$ adopts the innovation, so $x_i$ changes from 0 to 1. 
As this is a one-way process, unlike voter models \cite{castellano2009statistical}, the state at the next time-step is determined by:
\begin{equation}
x_i(t+1) = \left\{
  \begin{array}{l l}
    1 & \quad \text{if $x_i(t)=1$,}\\
    1 & \quad \text{if $x_i(t)=0$ and $u_i(t)>\theta_i$,}\\
    0 & \quad \text{otherwise,}\label{eqn:update}\\
  \end{array}\right.
\end{equation}
where $\theta_i$ is the threshold and $t$ is time.
The total utility $u_i$ is a combination of both personal and social benefit \cite{delre2010will}.
The personal benefit $p_i$ is a measure of the perceived intrinsic benefit to the individual of acquiring the innovation.
The social benefit can be split into both the direct influence from an individual's peer group and the (possibly indirect) influence from society in general \cite{valente1996social}, which could be affected by the desire to fit in with the social norm on different levels.
Thus our utility model  has three factors that can be given relative weightings $\alpha_i$, $\beta_i$ and $\gamma_i$, with $\alpha_i+\beta_i+\gamma_i=1$. 
The total utility is, therefore, given by:
 \begin{equation}
u_i(t) = \alpha_i p_i + \beta_i s_i(t) + \gamma_i m(t),\label{eqn:util}
\end{equation}
where $s_i$ is the average of $x$ within a node's \emph{network neighbourhood}:\\
\begin{align}
s_i(t) &= \frac{1}{k_i} \sum_j^N A_{ij} x_j(t),
\end{align}
and the \emph{degree} of node $i$ is $k_i=\sum_{j=1}^N A_{ij}$.
The mainstream social norm $m$ is the average of $x$ over the entire population:
\begin{align}
 m(t) &= \frac{1}{N}\sum_i^N x_i(t).
\end{align}
This can be made equivalent to previous threshold diffusion models such as \cite{watts2002simple} by setting $(\alpha,\beta,\gamma)=(0,1,0)$.
We define $m_0=m(t=0)$ to be the initial fraction of adopters.
For the rest of the discussion we make the simplifying assumption that  all nodes take the same values of $\alpha$, $\beta$, $\gamma$, $p$ and $\theta$, so that only the value of $s_i(t)$ differs between the nodes, which corresponds to the network effects. 

\subsection{Modelling social networks}

For these investigations we connected the nodes using several network models.
\emph{Erd\H{o}s--R\'enyi} random graphs \cite{erdos1960evolution}, where edges are connected with a probability $p_e$  between pairs of nodes, were investigated for their analytical simplicity, due to the lack of correlations between neighbourhoods of nodes, allowing probabilistic arguments to be used \cite{gleeson2012accuracy}.
The \emph{Watts--Strogatz} scheme \cite{watts1998collective}, where $n$-nearest-neighbour lattices have edges removed and reassigned according to a probability $p_r$, allowing interpolation between regular (possibly highly clustered) topologies and random networks, were also investigated, as they are frequently studied and provide a simple way to vary the clustering (\emph{transitivity}) of the system.

\emph{Community based} models, such as Newman's highly clustered networks \cite{newman2003properties}, can better represent real-world social interactions in a number of ways.
Here, links between nodes are assigned via their association with $G$ \emph{groups} each, which are randomly chosen from a set of $W$ groups in total. 
Association with two or more groups per node is necessary for the network to be fully connected and total transmission to be possible. 
Nodes are linked to $L$ other nodes within each group, so the total number of groups $W$ and resulting number of members of each group $M=GN/W$ provides a means of varying the clustering while keeping other network parameters fixed.
These networks also have locally higher clustering between members of a group than out to the rest of the network.
For these models it would also be possible to include some geographical information to the nodes and groups in the network (c.f.~Hamill and Gilbert's \emph{Social Circles} model \cite{hamill2009social}), as discussed further in \S \ref{sec:conc}.

Although widely used, we do not consider \emph{scale-free} networks, such as the \emph{Barab\'asi--Albert} (preferential attachment) model, here.
This is because, as well as blurring the numerical results and complicating the analytical treatment by introducing a distribution of the node degrees, most individuals are only  likely to  communicate with a small selection of their contacts about adoption of innovation. 

\section{Numerical methods}\label{sec:numerics}

\begin{figure}[h!]
\centering
(a)\includegraphics[width=0.45\linewidth]{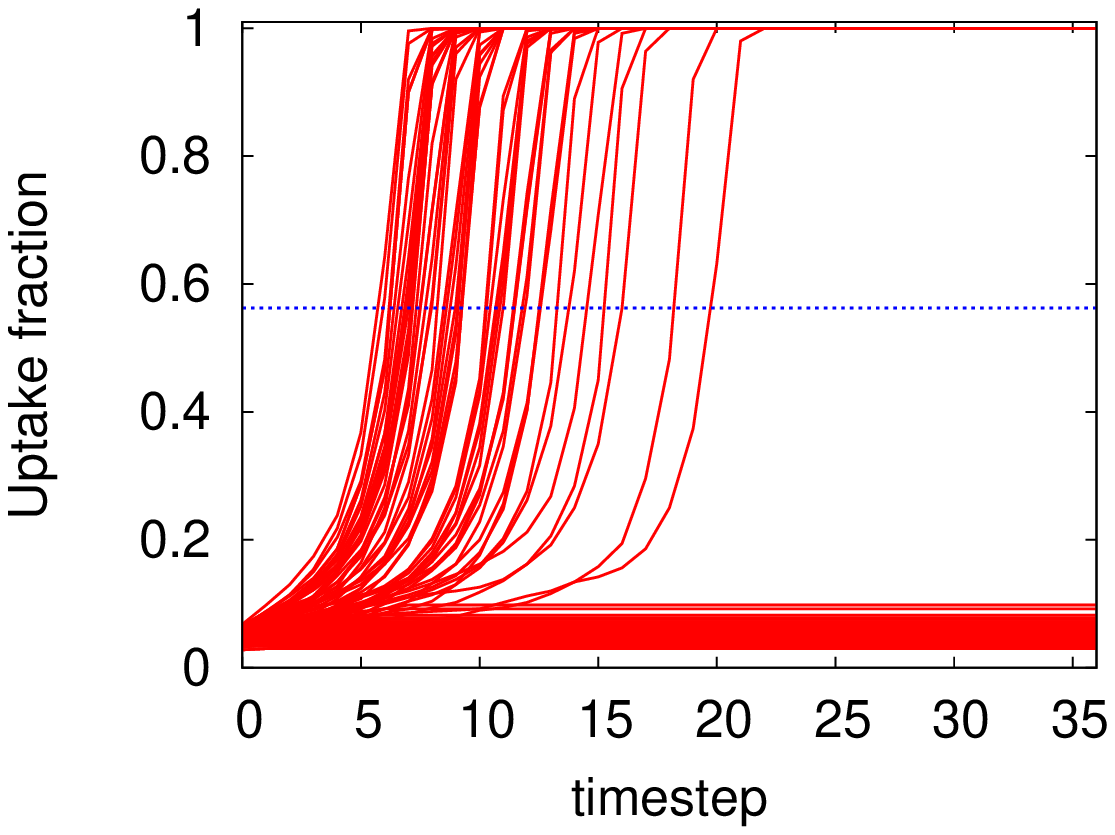}
(b)\includegraphics[width=0.45\linewidth]{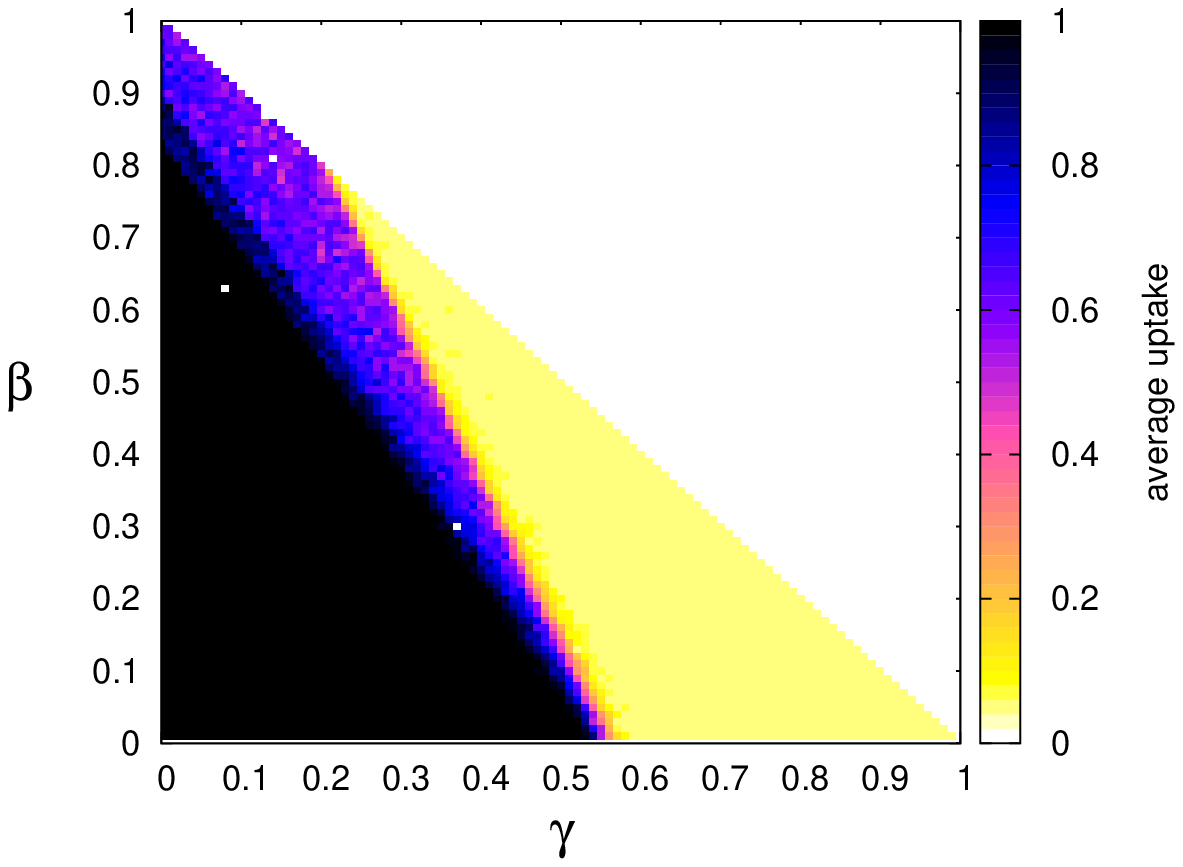}
\caption{Simulations of the adoption model on a community based network, with $N=500$ nodes linked to others via their shared association with  groups.  Each node is randomly assigned to $G=2$ groups, from a total of $W=100$, and is linked with $L=5$ other nodes per group. (a) Runs starting from 100 realisations of a 5\% level of initial adoption ($m_0=0.05$), with model parameters $\alpha=0.05, \beta=0.8, \gamma=0.15, p=0.5, \theta=0.25$. The mean uptake over the 100 realisations at time-step 36 is 56\%. (b) The mean uptake at values in the $\beta,\gamma$ parameter space, with $\alpha=1-\beta-\gamma$. All nodes take the same parameter values as each other.}\label{fig:example}
\end{figure}

Given a particular  network,  initial set of adopters and parameter values ($\alpha,\beta,\gamma,p,\theta$), we ask whether or not the innovation \emph{succeeds}, that is, do most individuals adopt the innovation.
Aside from the challenges inherent in modelling the complex factors affecting behaviour, as well as in quantifying models from social data, dynamical processes on complex networks themselves often exhibit  dependence on the starting conditions, such as which nodes are chosen as the initial seed.
To illustrate this, the simulation results for a particular network and a single choice of parameters are given in Figure \ref{fig:example} (a), showing the uptake over 36 time steps for 100 random realisations of the initial seed.
In each case, 5\% of the nodes are chosen as \emph{seed nodes} at random and are set to 1, with the rest  initialised at 0.
It can clearly be seen that while many runs lead to successful adoption, defined as most or all nodes becoming active, there is a clear subset of about 44\% of initial conditions that stagnates and does not take over an appreciable proportion of the network.

Given this dependence, the ensemble average uptake over all 100 realisations of the choice of seed-nodes is computed for each chosen set of parameter values. 
The detailed structure of the network is randomised at each set of parameter values.
In order to look at the influence of the parameters on the expected uptake, this calculation is repeated over a range of parameter values and plotted in the $\beta, \gamma$ parameter space, with $0 \leqslant \beta \leqslant 1$, $0 \leqslant \gamma \leqslant 1$ and $\alpha=1-\beta-\gamma$.
An example plot is shown in Figure \ref{fig:example} (b).
Two lines can be seen dividing regions of the parameter space where adoption is total (black), where uptake is about 65\% (blue) and where it stagnates at levels up to only slightly above the initial seeding level of 5\% (yellow).

\section{Results and analysis}\label{sec:anal}     

The procedure described in the previous section was carried out for different network topologies in order to investigate the general behaviour of the model.
The results all show distinct regions of either near total success or stagnation, with a few exceptions where there is a region of intermediate probability of success. 
In this section we present the results of these simulations along with an analysis into the origin of the dividing lines between these regions in  parameter space.      
This is presented first in terms of  Erd\H{o}s--R\'enyi random (ER) networks, and later for other network models.

\subsection{Number of neighbours needed to induce uptake}

In order to  understand which parameter values are likely to lead to a higher level of adoption, it is important to consider the properties of the nodes' local network environment.
Each node experiences the same $p$ and $m$ in Equation (\ref{eqn:util}), so the only difference between the nodes is $s_i$, the fraction of its neighbours that are active.
For any set of parameter values and current adoption level $m(t)$, there is a critical fraction $s^*(t)$, defined by:
\begin{equation}
s^*(t) = \dfrac{\theta - {\alpha} { p} - {\gamma} {m}(t)}{\beta}  \label{eqn:rule2},
\end{equation}
such that if $s_i \geqslant s^*$, the node will adopt the innovation in the next time-step.
The required number of active contacts $Y_i^*$ (an integer) is given by:
\begin{equation}
Y_i^*(t) = \left\lceil k_i s_i^*(t)\right\rceil,\label{eqn:threshnum}
\end{equation}
where $\left\lceil \bullet \right\rceil$ denotes the smallest integer $\geqslant\bullet$.
If the actual number of active contacts
\begin{equation}
Y_i = \sum_j A_{ij} x_j 
\end{equation}
is at least $Y_i^*$, then node $i$ will take up the innovation in the next time-step.
To simplify matters, we consider first networks where the nodes all have the same (or similar) degree $\bar k$, defined by:
\begin{equation}
\bar k = \frac{1}{N} \sum_{ij} A_{ij}.
\end{equation}
In this case we can define the critical number of active contacts to be
\begin{equation}
 Y^* = \left \lceil  \bar k s^*(t) \right \rceil = \left \lceil \bar k \left(\dfrac{\theta - \alpha p - \gamma m(t)}{\beta}\right) \right \rceil,\label{eqn:actnum}
\end{equation}
and the success or otherwise of the innovation uptake across the network can be understood in terms of $Y^*$.
If $Y^*=0$, then all nodes immediately adopt the innovation.
If $Y^*=1$, then any node connected to one of the initial seed nodes adopts the innovation, and the innovation uptake will be successful in a fully connected network.
If $Y^*=2$, then a node must be connected to two active nodes in order to adopt.
We show below that for reasonable values of $m_0$, and for a wide class of networks, this leads to success.
The case $Y^*=3$ will also be discussed below.

\begin{figure}[h!]
\centering
(a)\includegraphics[width=0.45\linewidth]{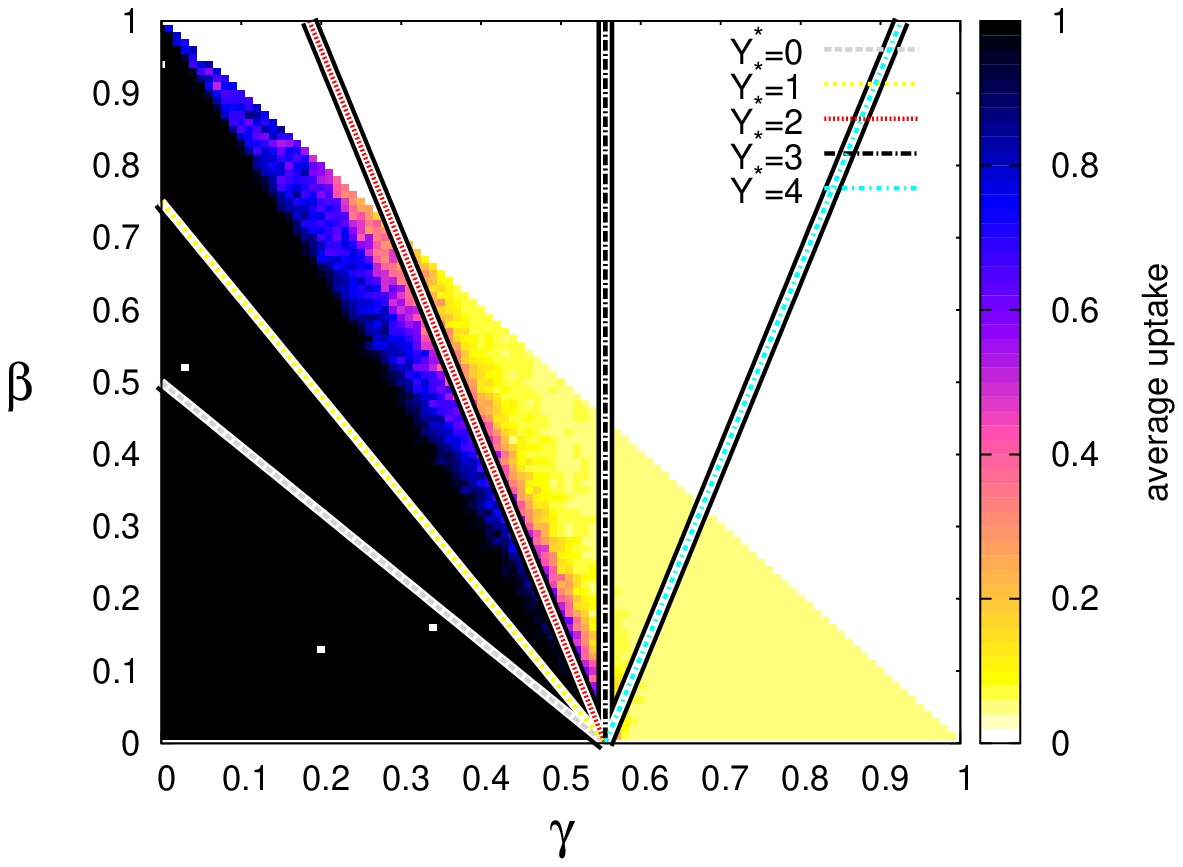}
(b)\includegraphics[width=0.45\linewidth]{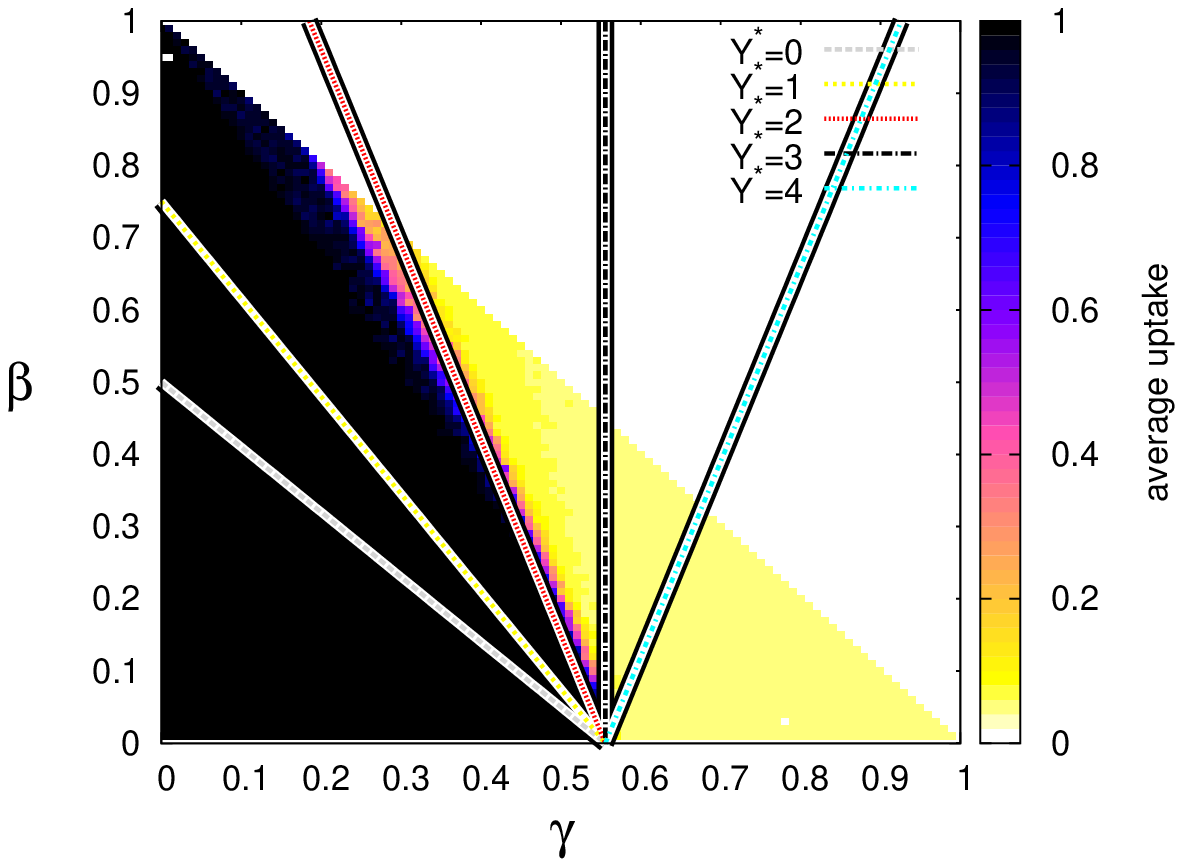}
(c)\includegraphics[width=0.45\linewidth]{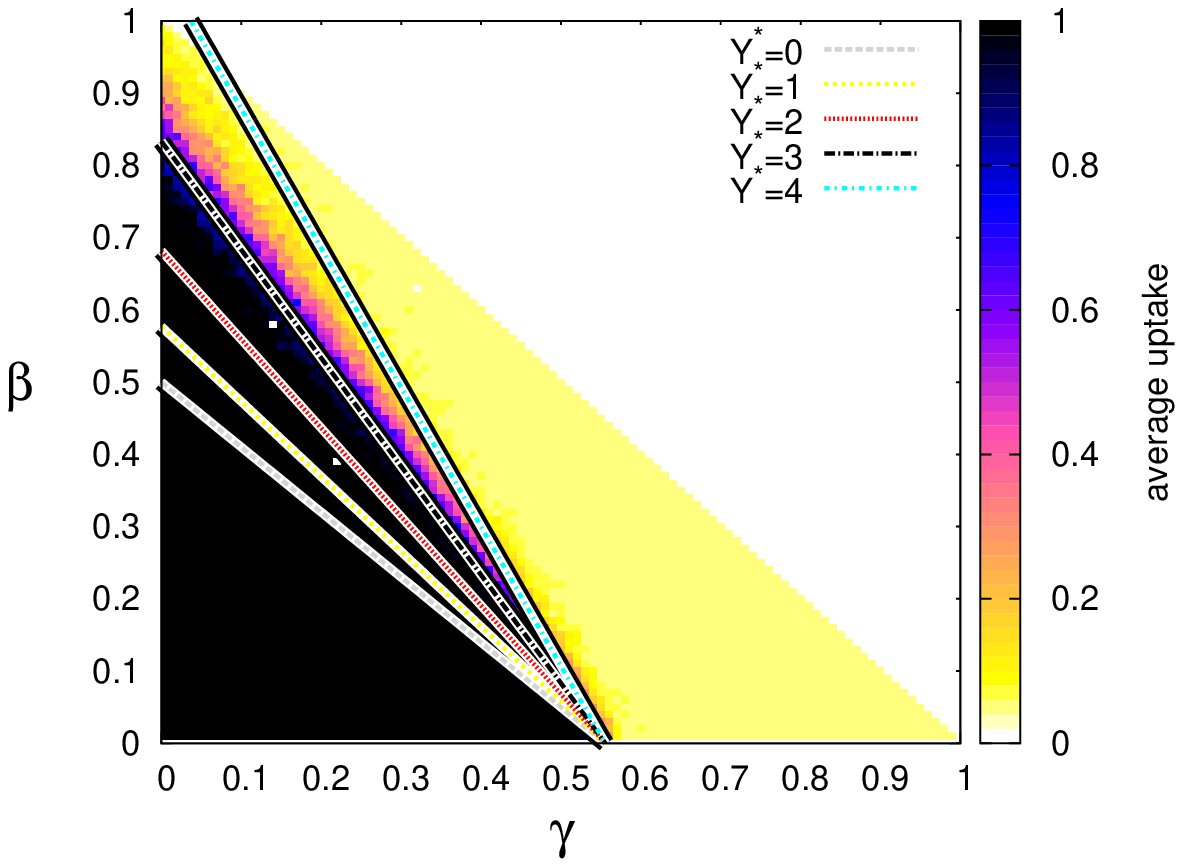}
\caption{Comparison of simulations of the adoption model on random Erd\H{o}s--R\'enyi networks.  As in all cases $\theta=0.25$ and $p=0.5$. The mean node degree $\bar k$ and network size $N$ values are: (a) $\bar k=6$, $N=500$; (b) $\bar k=6$,  $N=2000$; (c) $\bar k=15$, $N=500$. The results of the analysis in \S \ref{sec:anal} are also plotted for $Y^*=0,1,2,3,4$.  The lines of $Y^*$ in the ($\beta,\gamma$) plane show the upper limits of ($\beta,\gamma$) for each value of the critical number of neighbours $Y^*$.  For example, two active neighbours are required to trigger adoption to the left of the $Y^*=2$ line but this changes to three on the right side of the line.}\label{fig:plots1}
\end{figure}

Using $\alpha=1-\beta-\gamma$, Equation (\ref{eqn:actnum}) can be inverted to give:
\begin{equation}
\beta = \bar k \left(\dfrac{\gamma(m(t) - p) + p - \theta}{\bar k p - Y^*}\right).
\end{equation}
This gives, for different $m(t)$, and for different integers $Y^*$, a sequence of lines in the ($\beta,\gamma$) parameter plane.
These lines (with $m(t)=m_0$) are overlaid on numerical results in Figure \ref{fig:plots1} for random (ER) networks of different size ($N=500$, $N=2000$) and different average degree ($\bar k = 6$, $\bar k = 15$).
In Figure \ref{fig:plots1}(a) ($N=500$, $\bar k = 6$), uptake is successful for most values of ($\beta,\gamma$) to the left of the $Y^*=2$ line.
This boundary becomes sharper for $N=2000$ and agrees more closely with the theoretical $Y^*=2$ line in Figure \ref{fig:plots1}(b).
For $\bar k = 15$, $N=500$ (Fig.~\ref{fig:plots1}(c)), it is the $Y^*=3$ line that separates success from failure.

\subsection{Probability of induced uptake in random networks}\label{sec:anal3}

We begin by calculating the probability that the condition $Y_i\geqslant Y^*$ is satisfied.
For random networks with independent random assignment of initial conditions, where  all nodes have statistically homogeneous local network environment and the same node degree $k$, this is:
\begin{equation}
 P(Y \geq Y^*) = \sum_{n=Y^*}^{k} \binom{k}{n} m^{n} (1-m)^{(k - n)}.\label{eqn:prob1}
\end{equation}
This is just the probability of at least $Y^*$ nodes being adopters, from $k$ nodes in the neighbourhood, given that each node has a probability $m$ of being an adopter.
Clearly $P(Y\geq Y^*)$ is zero if $Y^* > k$.

For the specific examples in Figure \ref{fig:plots1} the outcomes in the different sectors of the parameter space can be understood using these ideas.
If we define $Z$ to be the number of nodes in the whole network whose neighbourhood influences them to adopt, then the probability that at least one further node will be influenced somewhere in the network is $P(Z\geq1)=1-P(Z=0)$, where the probability that no neighbourhoods are influential is $P(Z=0)=\left(P(Y<Y^*)\right)^N$, and $P(Y<Y^*)=1-P(Y\geqslant Y^*)$ which, using (\ref{eqn:prob1}) gives the expression:
\begin{equation}
 P(Z\geq1) = 1-\left(1-\sum_{n=Y^*}^{k} \binom{k}{n} m^{n} (1-m)^{(k - n)}\right)^N,\label{eqn:PZ1}
\end{equation}
which, for small $m$ approximates to:
\begin{equation}
P(Z\geq1) \approx 1-\left(1-\binom{k}{Y^*} m^{Y^*}\right)^N.\label{eqn:PZapp}
\end{equation}
For $k=6$, $N=500$ and $m_0=0.05$ this gives $P(Z\geq1)\approx 1$ for both $Y^*=1,2$, but for $Y^*=3$, $P(Z\geq1)\approx 0.7$.
For the $Y^*=2$ case, the probability of a site having an influential neighbourhood is $P(Y\geq Y^*)\approx0.0375$ and the expected number of nodes satisfying this condition is this probability multiplied by the number of remaining (inactive) nodes $N(1-m)P(Y\geq Y^*)\approx 18$, doubling adoption at each time-step.
However, in the $Y^*=3$ sector, the probability of any chosen node having the required number of neighbours is $P(Y\geq Y^*)\approx 0.0025$, resulting in only one expected new active node with a network size $N=500$, easily leading to stagnation.
The $Y^*=2$ line becomes a sharper dividing line for larger networks (Fig.~\ref{fig:plots1} (b)), as finite size fluctuation effects become less important.
The value of $Y^*$ which is the dividing line changes for $\bar k=15$ (Fig.~\ref{fig:plots1} (c)), where $N(1-m)P(Y\geq Y^*=3)\approx 27$, again of the order of the initial seeding level, and so adoption is successful for $Y^*=3$, but fails when $Y^*=4$.

\subsection{Propagation on random networks}

In Equation (\ref{eqn:actnum}), $Y^*$ is a decreasing function of time, since $m(t)$ can only increase.
As argued above, if $Y^*=0$ or 1, uptake will be successful.
In this section, we estimate the likelihood that $Y^*$ will decrease from a larger initial value to 1, and how long this will take.

The expected number of nodes that will become adopters in this time-step is the number of nodes that have not adopted the technology, $N(1-m)$, multiplied by this probability.
The overall fraction of active nodes will therefore increase by:
\begin{equation}
\Delta m = (1-m) P(Y \geq Y^*),\label{eqn:dm}
\end{equation}
giving the updated value of $m(t+1)$ as a function of the current value $f(m(t))$:
\begin{equation}
     m(t+1) = m(t)+(1-m(t))P(Y \geq Y^*) \equiv f(m(t)).\label{eqn:fm}
\end{equation}
$Y^*$ is also a function of $m$ but, being an integer, can only take discrete values, so $f(m)$ is discontinuous when $Y^*$ changes.

\begin{figure}[h!]
\centering
(a)\includegraphics[width=0.45\linewidth]{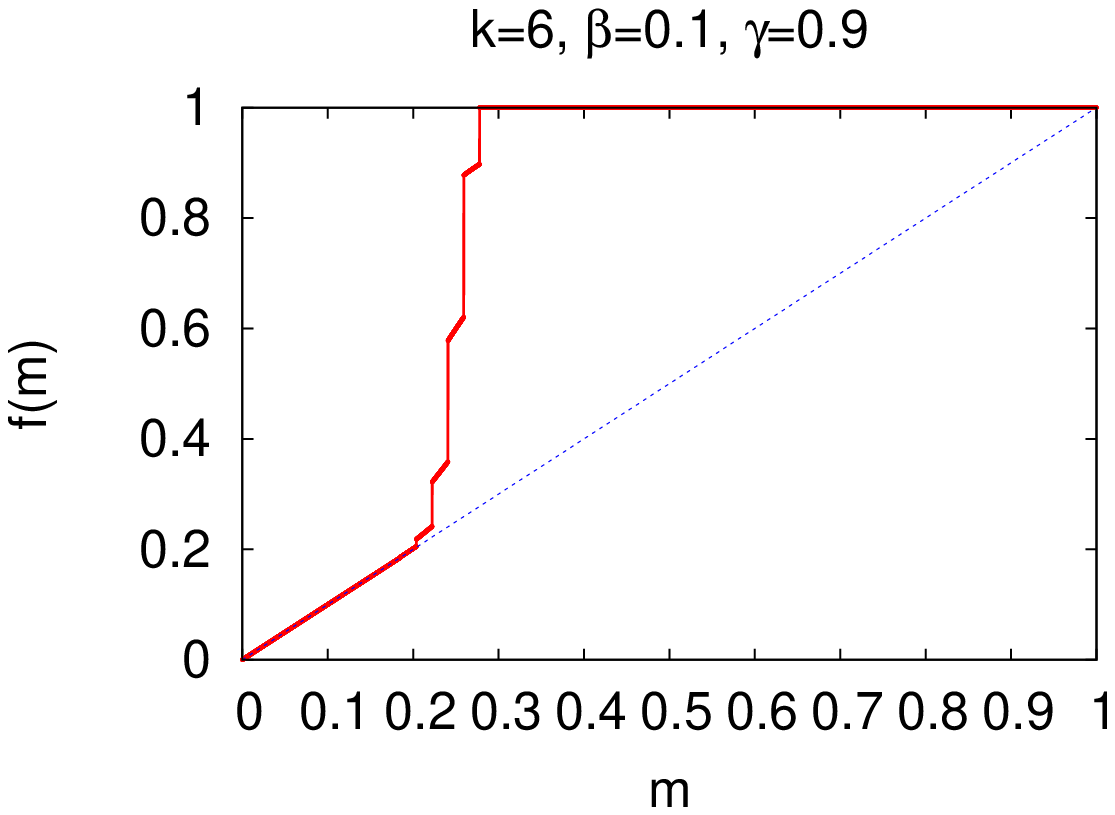}
(b)\includegraphics[width=0.45\linewidth]{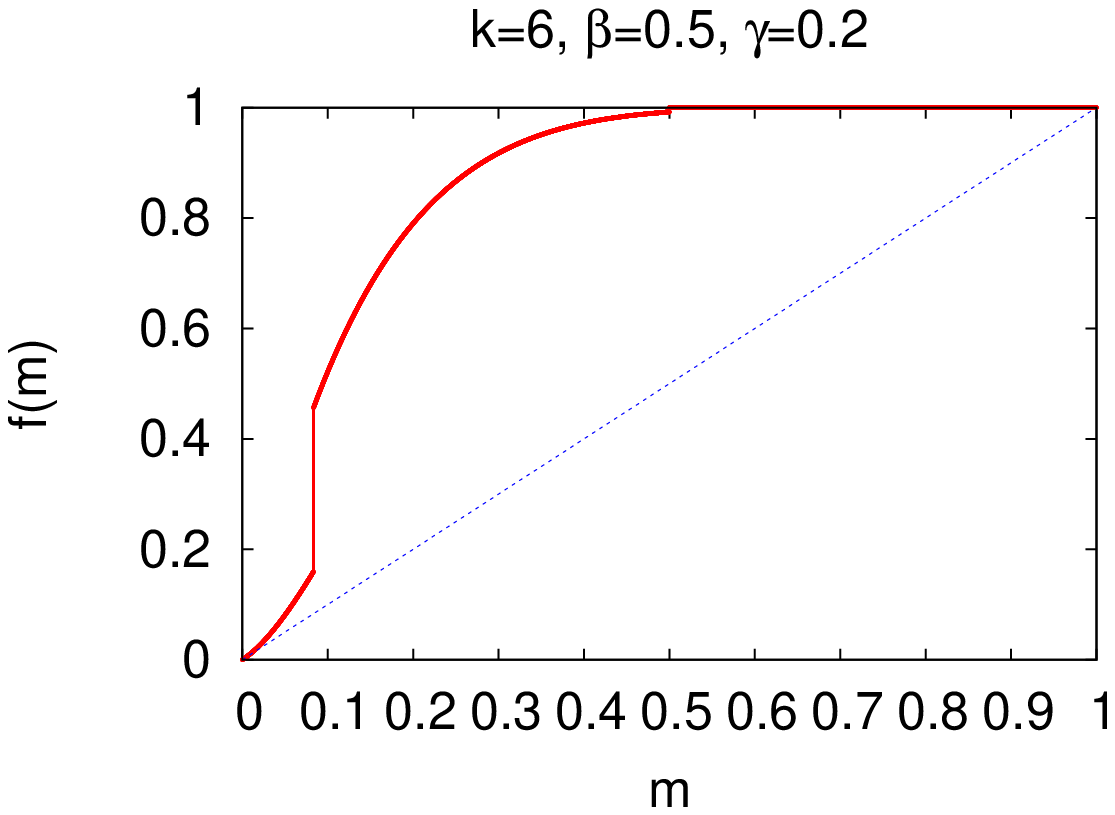}
(c)\includegraphics[width=0.45\linewidth]{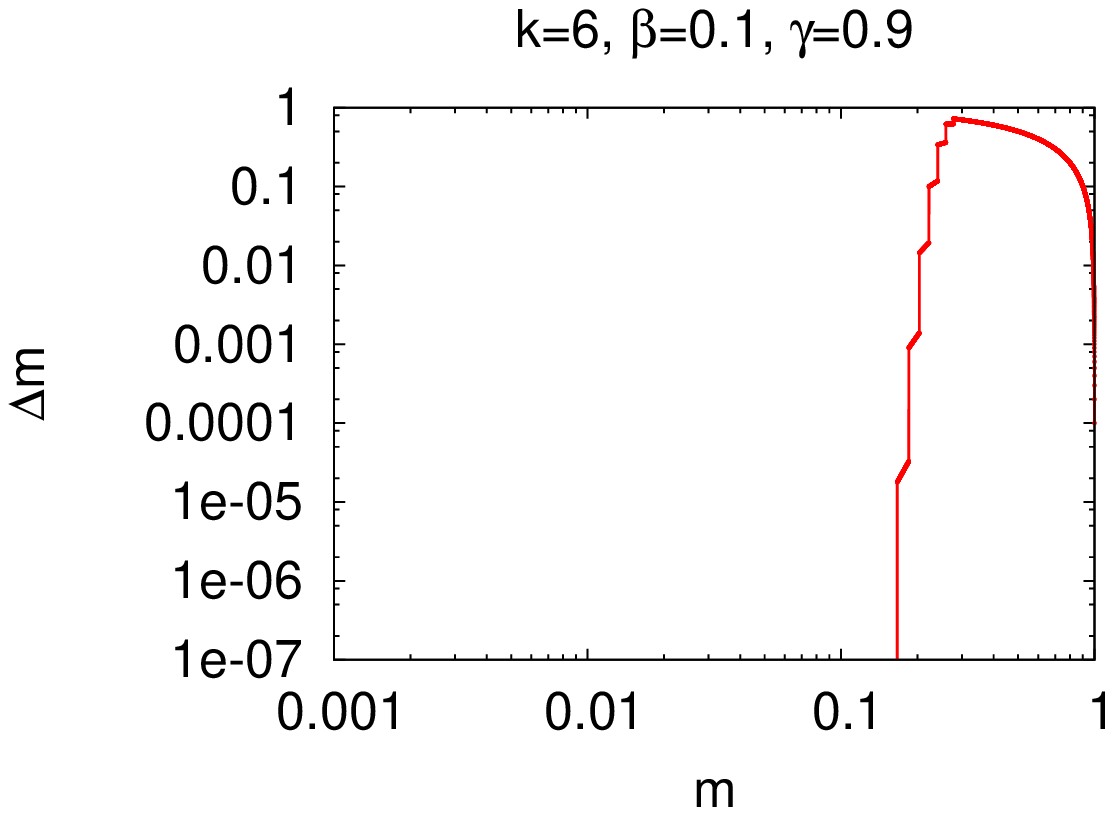}
(d)\includegraphics[width=0.45\linewidth]{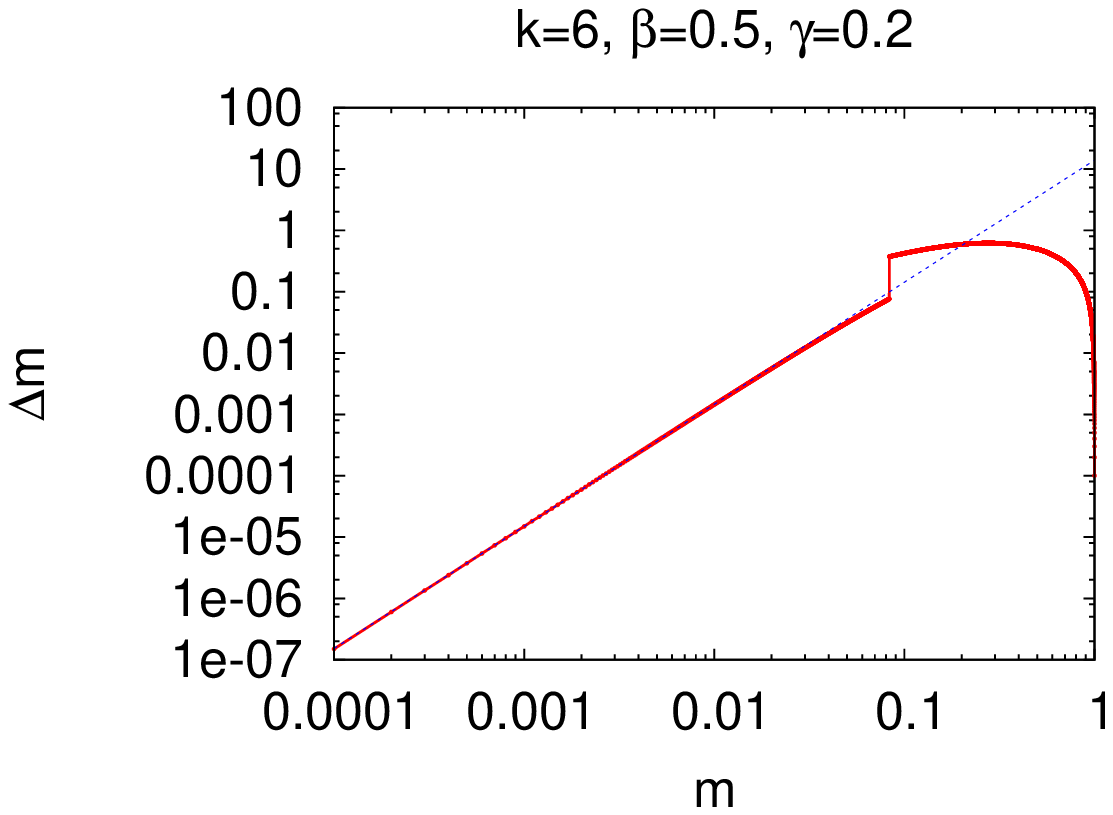}
\caption{Plots of Eqn.\ (\ref{eqn:fm}), for $k=6$, (a) $\beta=0.1, \gamma=0.9$ and (b) $\beta=0.5, \gamma=0.2$, and Eqn.\ (\ref{eqn:dm}) at the same values in (c) and (d) (on a log scale).  In (a) and (c) $Y^*>k$ for small $m$ so there is no further uptake, and in (b) and (d) $Y^*=2$ for small $m$ so uptake is disproportionately slow for small levels of seeding.  The diagonal is drawn for comparison in (a) and (b), while in (d) the fit to the initial slope is given.}\label{fig:rates}
\end{figure}

This gives a macroscopic, system level, version of the microscopic (individual) dynamics given by Equation (\ref{eqn:update}).
The functions in (\ref{eqn:fm}) and (\ref{eqn:dm}) are shown in Figure \ref{fig:rates} for some example parameter values.
In Figure \ref{fig:rates} (a) and (c) we have $Y^*>k=6$ for $m<0.1$, so $\Delta m=0$ in this range, and starting at an initial seed level of $m_0=0.05$ will lead to stagnation.
If $m_0 \geqslant 0.2$, $\Delta m$ is quite large and success will come quickly.
In contrast, in Figure \ref{fig:rates} (b) and (d), $Y^*=2$ for small $m$, and uptake can only happen if the network size $N$ is large enough to allow further nodes to be influenced.

For small values of the initial adoption $m \ll 1$ the following approximation can be made:
\begin{equation}
P(Y \geq Y^*) \approx \Delta m \approx  \binom{k}{Y^*} m^{Y^*},
\end{equation}
with $Y^*\approx\left\lceil \bar k \left( \dfrac{\theta - \alpha p}{\beta} \right) \right\rceil$ for small enough $m$.
If $Y^* > 1$ then the super-linear power of $m$ results in very small returns for small initial seeding, and it takes a time on the order of $m_0^{1-Y^*}$ for a successful outcome, for small $m_0$.

\subsection{Structured networks}

\begin{figure}[h!!]
\centering
(a)\includegraphics[width=0.45\linewidth]{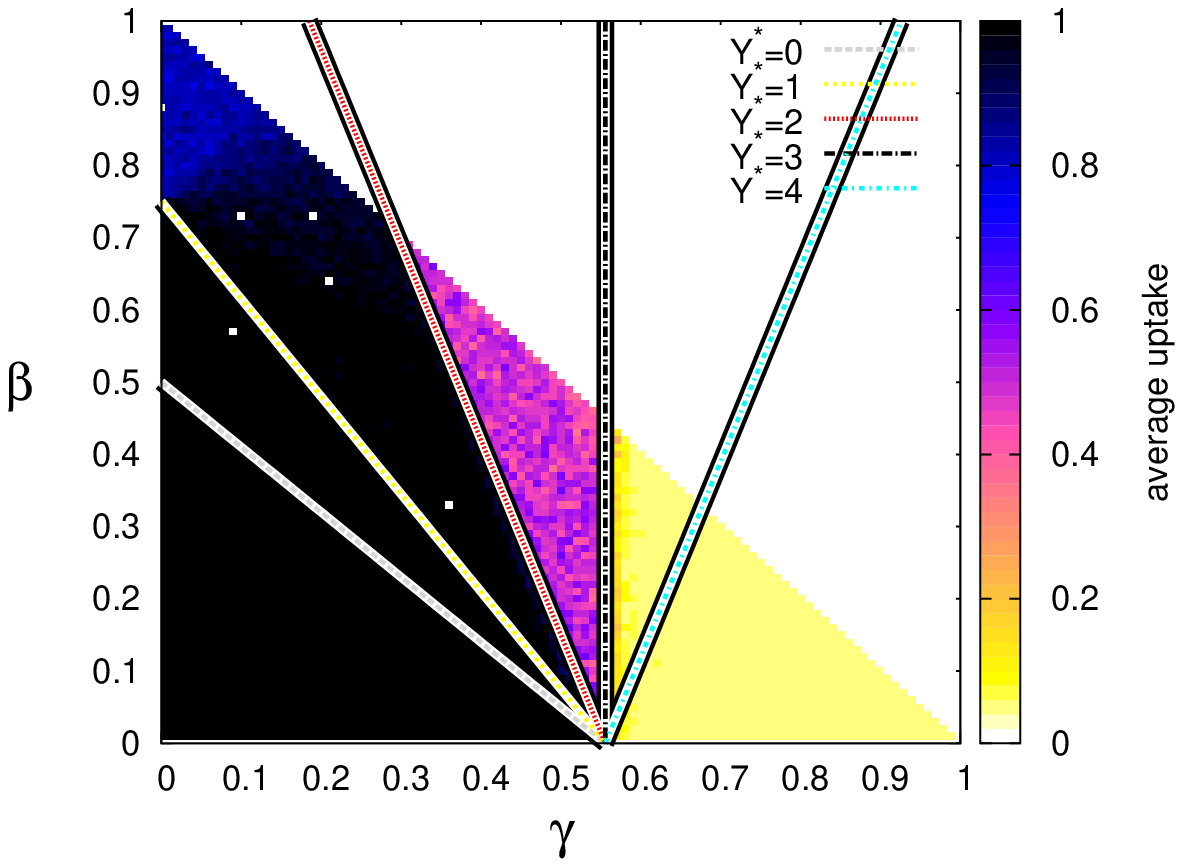}
(b)\includegraphics[width=0.45\linewidth]{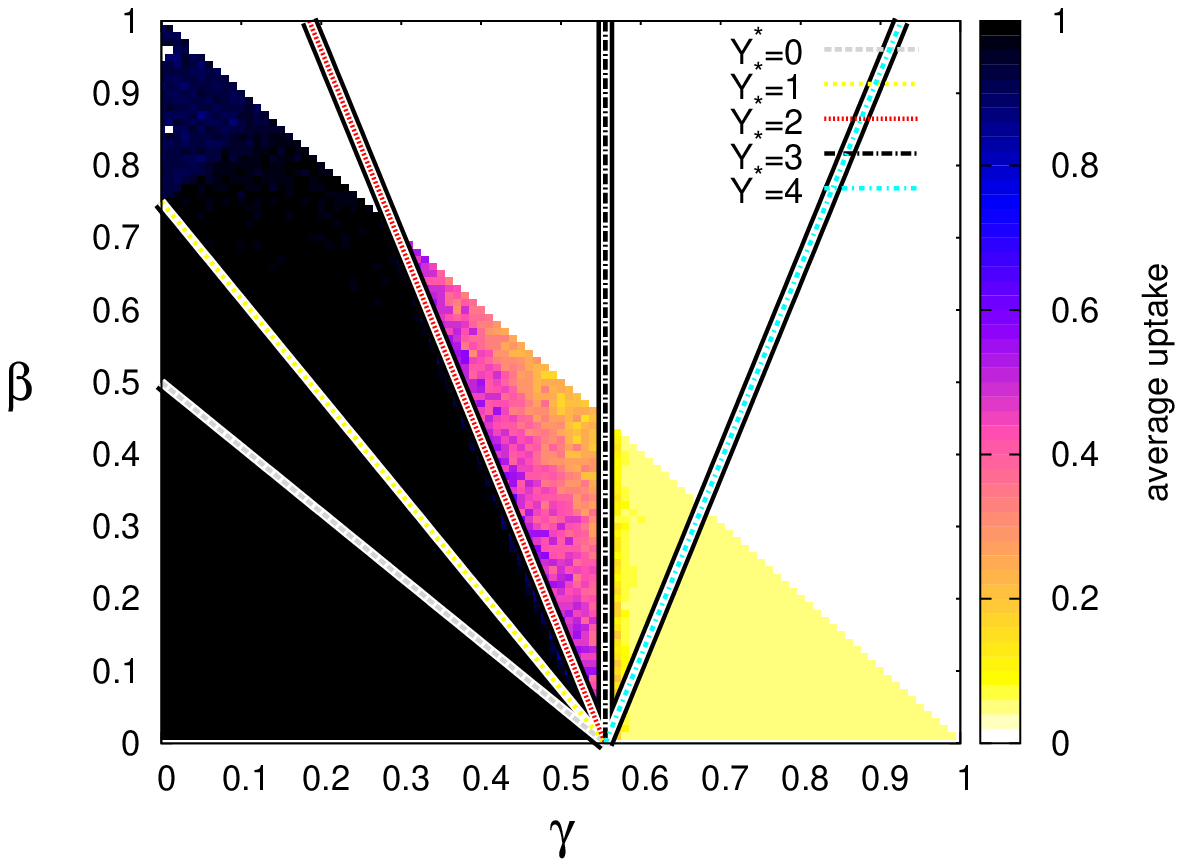}
(c)\includegraphics[width=0.45\linewidth]{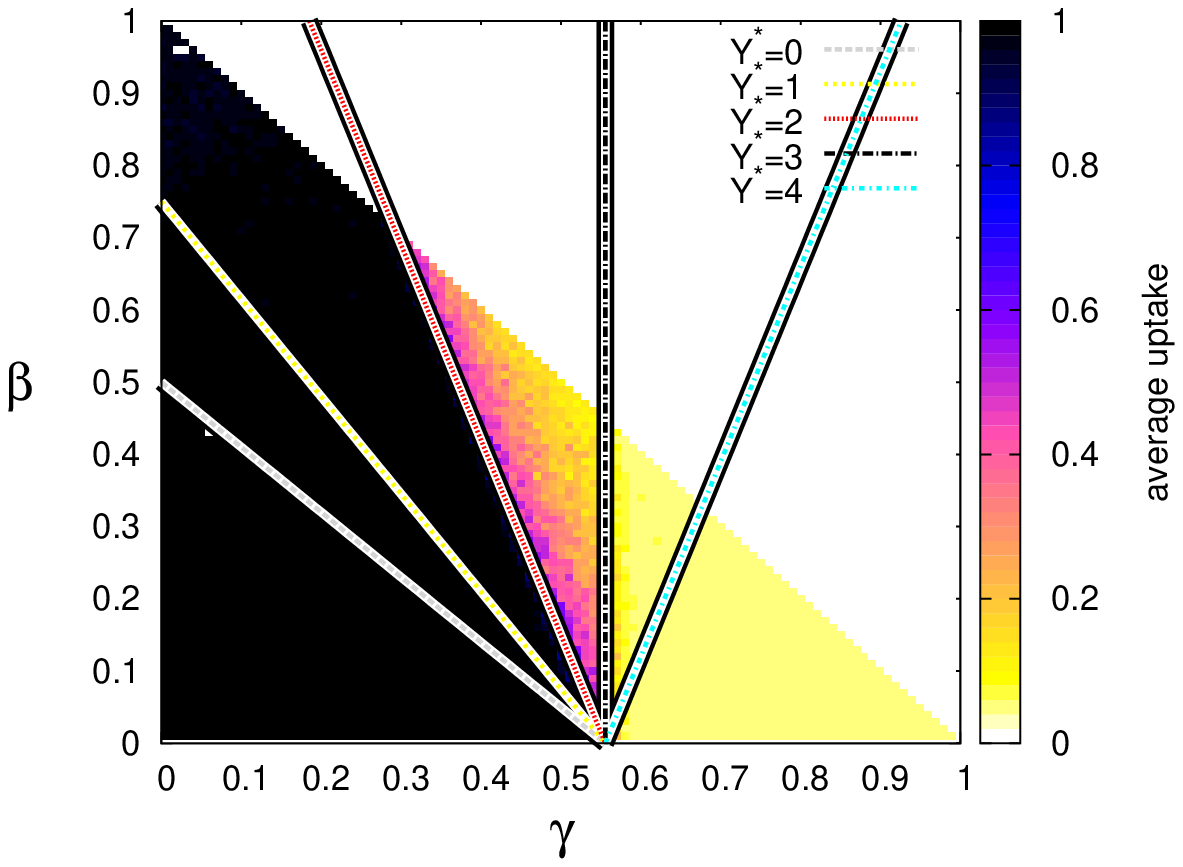}
(d)\includegraphics[width=0.45\linewidth]{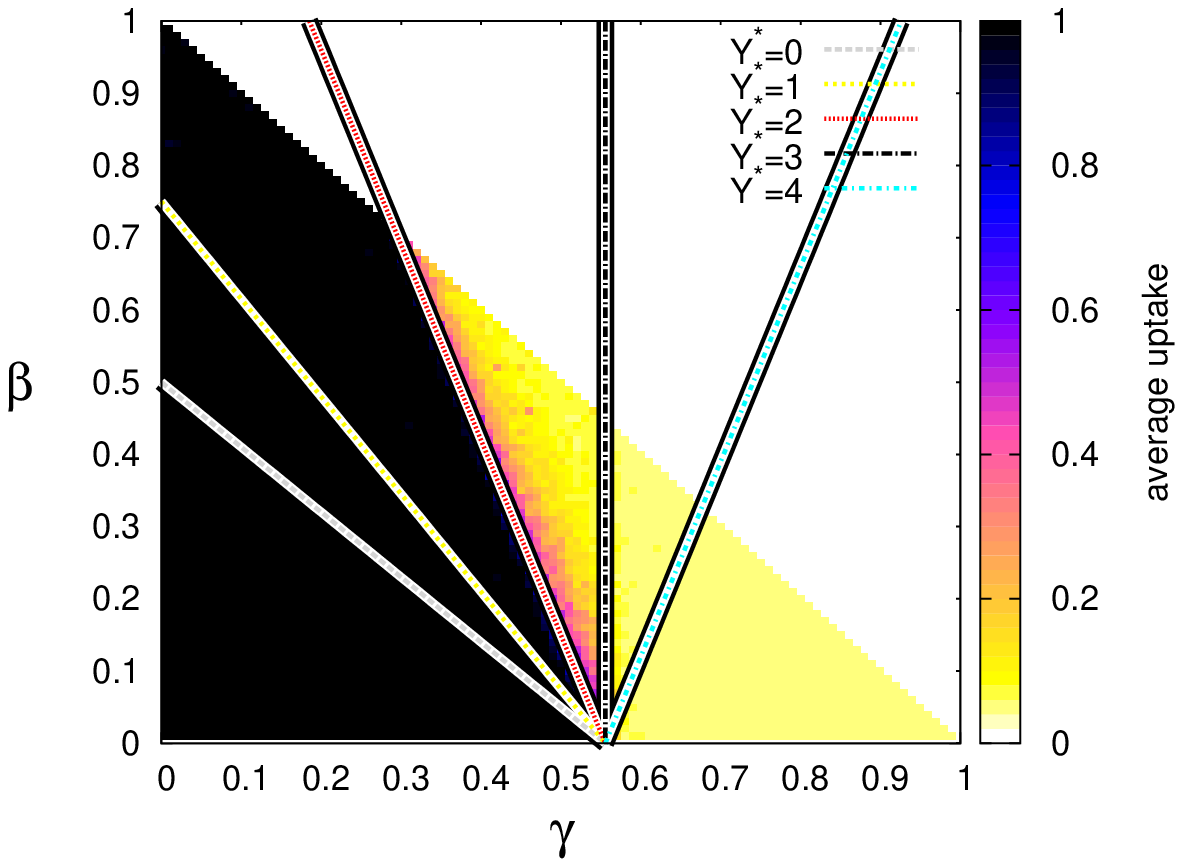}
\caption{Results on Watts--Strogatz networks with nodes initially connected to their six nearest neighbours on a one dimensional lattice and different rewiring probabilities $p_r$, resulting in different transitivity $c$: 
(a) $p_r = 0$, $c=0.6$; (b) $p_r=0.01$, $c=0.57$; (c) $p_r=0.02$, $c=0.53$; (d) $p_r=0.05$, $c=0.45$.  Model parameters are $p=0.5$ and $\theta=0.25$, with $N=500$.  Edges were swapped in pairs to preserve the node degree $k=6$.}\label{fig:plotsN}
\end{figure}

The approach above works for random networks, where independence of neighbourhoods can be assumed, and for randomly distributed initial seeds.
For networks with high clustering, where the clustering coefficient or \emph{transitivity}, defined as
\begin{equation}
c=\left(\dfrac{\text{\#(length 2 paths)}}{3\times \text{\#(triangles)}}\right),
\end{equation}
is close to 1, this is not the case.
This difference is demonstrated in the numerical simulations shown in Figure \ref{fig:plotsN}.
For these results the Watts--Strogatz scheme was used to rewire a highly clustered one dimensional lattice, where nodes are initially laid out in a ring and linked to their six nearest neighbours.
Due to the large number of triangles, this network has a high transitivity $c=0.6$, indicating a 60\% probability that any two neighbours of a node are also neighbours of each other.
For comparison, a completely random (ER) network of this size and edge density would have $c=0.01$.

The probability of rewiring is $p_r$, and for small values of $p_r$ where there is still significant clustering, the results show successful uptake in around 50\% of cases in the $Y^*=3$ sector, with either successful or stagnated adoption for $Y^*$ values below or above this, respectively.
This behaviour for $Y^*=3$ would not be expected in random networks at these values, where the random probability of any particular neighbourhood triggering an adoption $P(Y\geq Y^*) \approx 0.002$, resulting in only a single expected further uptake per realisation.
This would result in very slow uptake or stagnation on finite networks, which is indeed what happens for large $p_r$, where $c$ is small and the network is again more random in structure (Fig.~\ref{fig:plotsN} (d)).

\begin{figure}[b!]
\centering
(a)\includegraphics[width=0.4\linewidth]{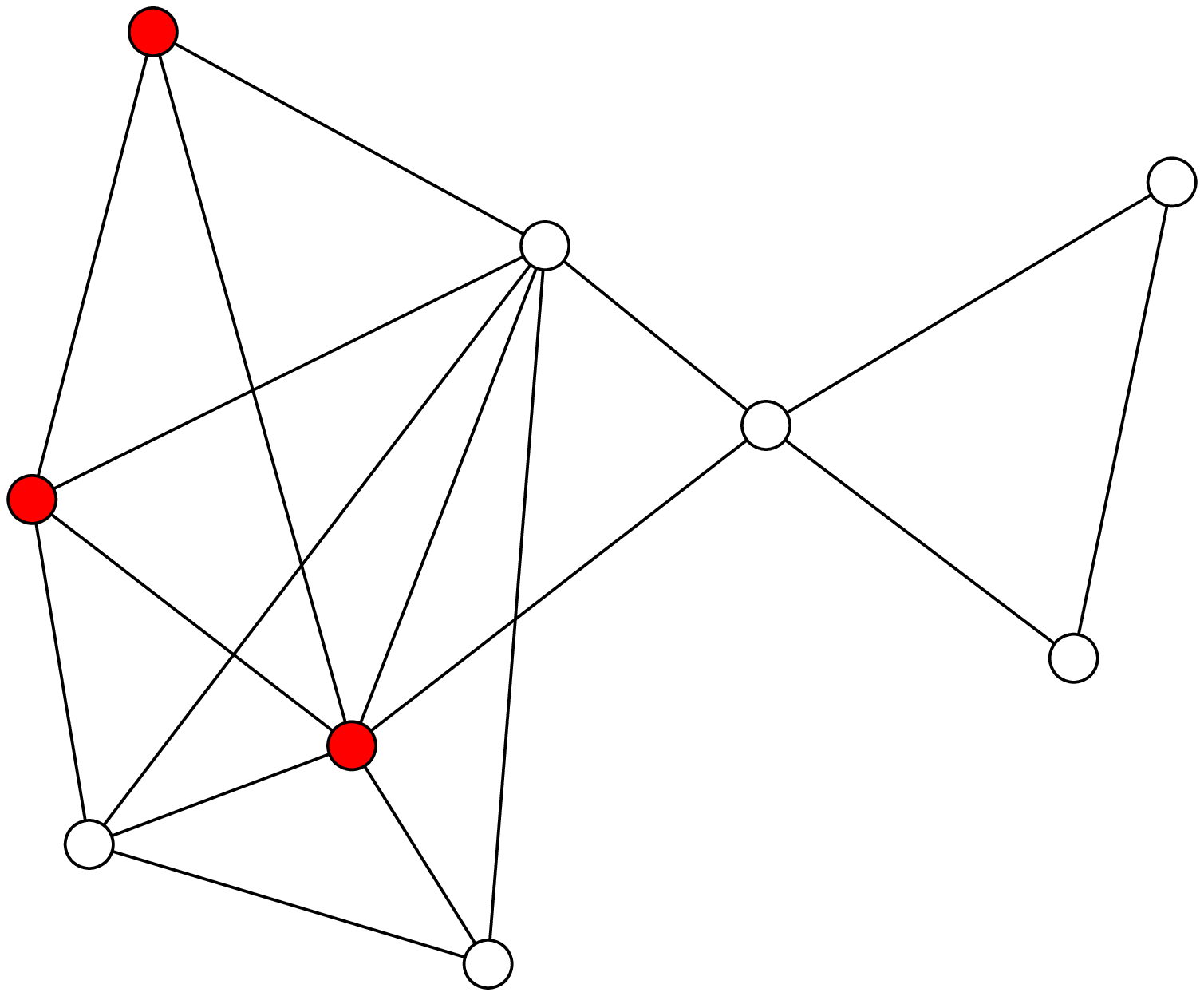}
(b)\includegraphics[width=0.4\linewidth]{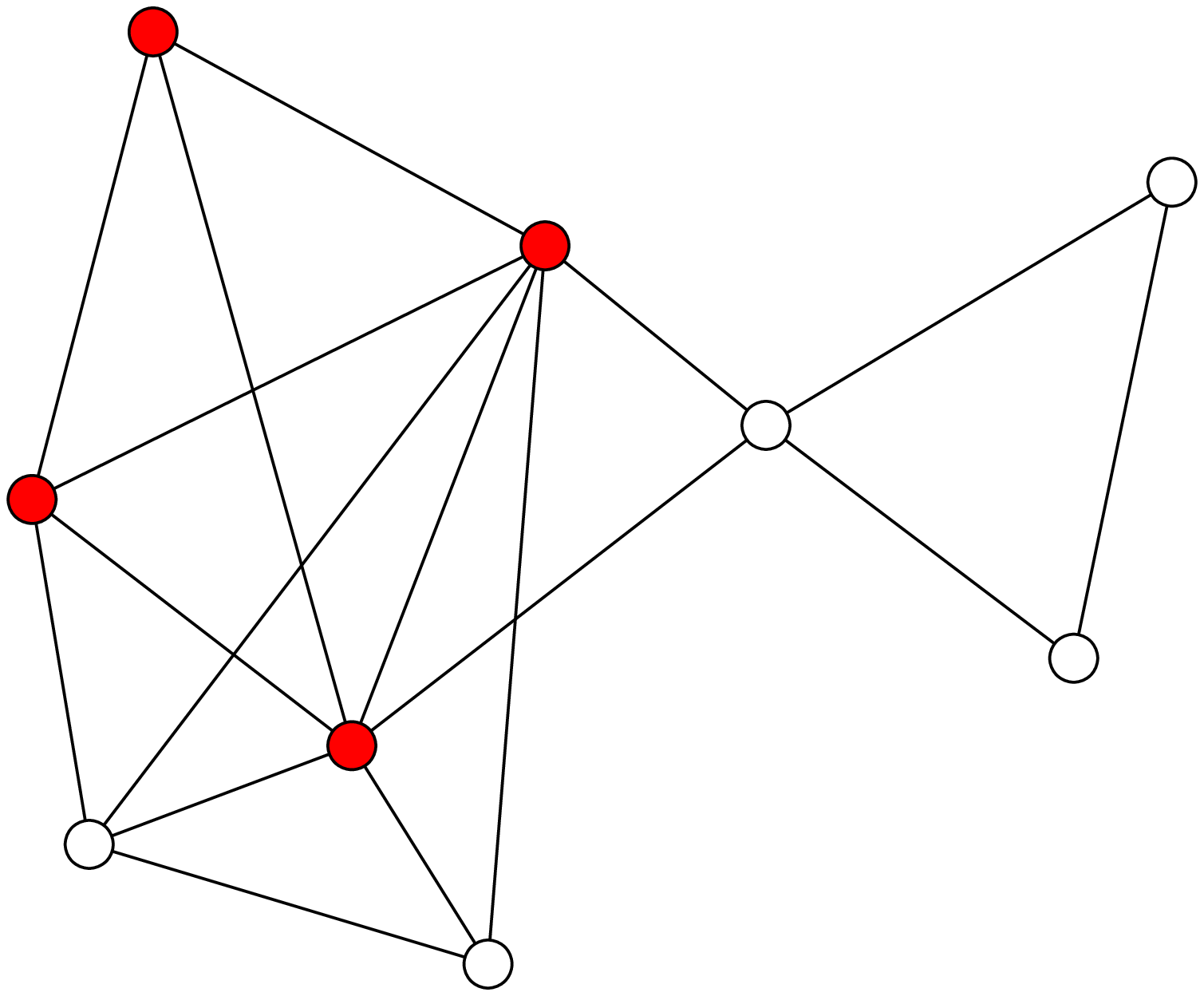}
(c)\includegraphics[width=0.4\linewidth]{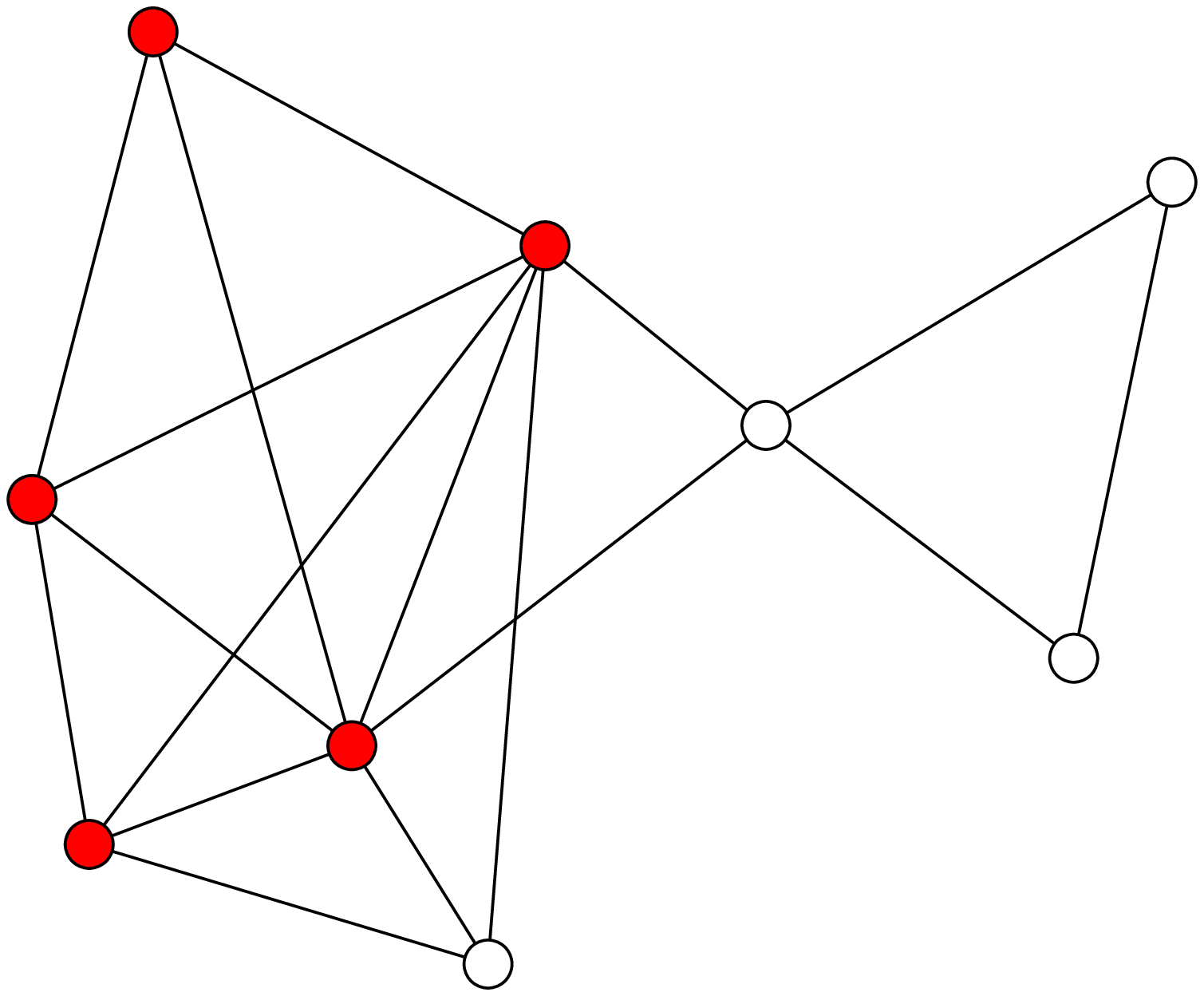}
(d)\includegraphics[width=0.4\linewidth]{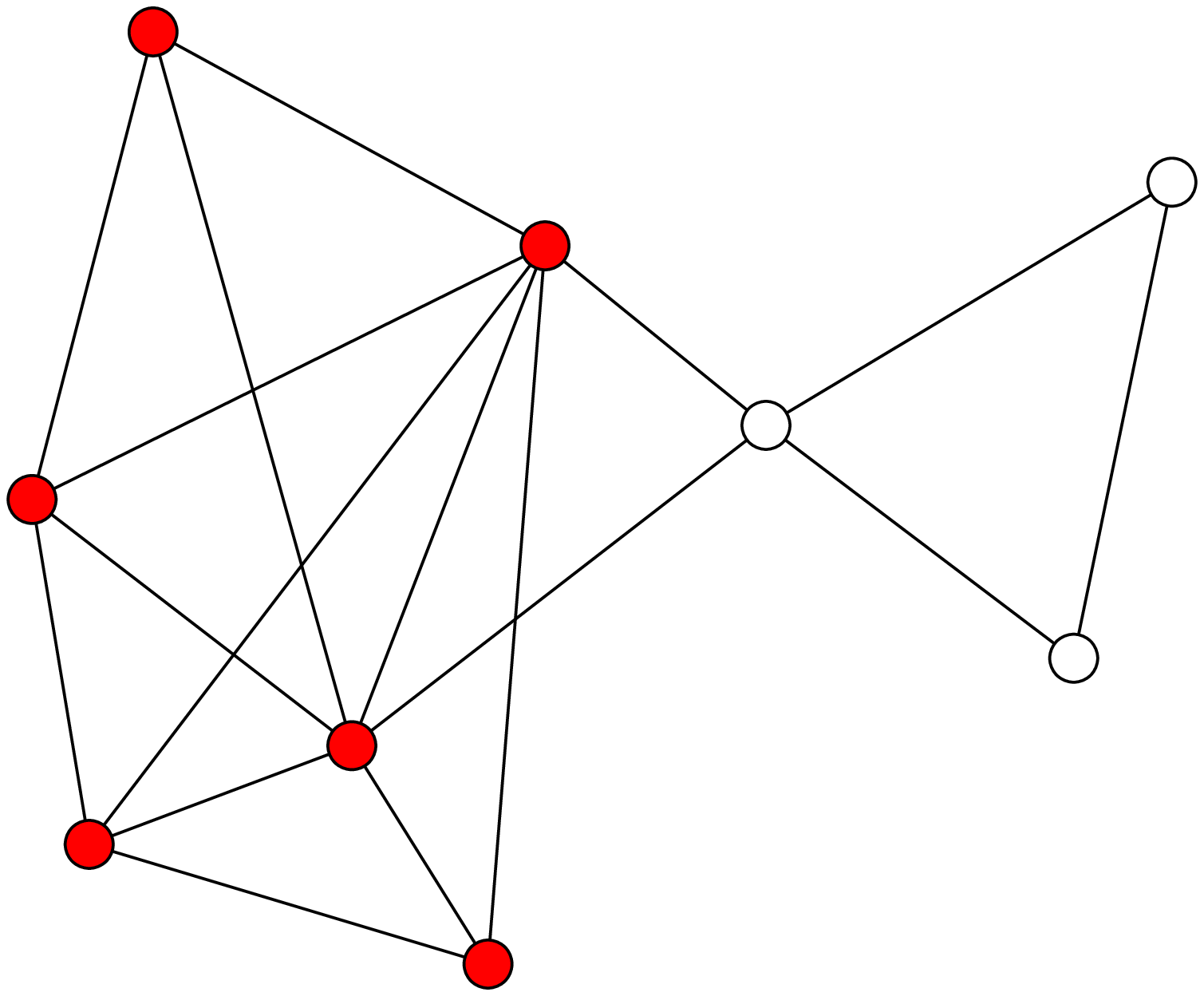}
 \caption{An example of uptake propagation via clustering, in a network where nodes require three active neighbours to become active themselves. The sequence of uptakes is shown in (a)--(d), starting from a single node influenced by three of its neighbours, which then induces further uptake via 3-member overlap of 4-cliques.}\label{fig:clique}
\end{figure}

The enhanced success of uptake in these cases is due to the clustered structure favouring continuing propagation from any influential neighbourhood. 
In clustered networks, any neighbour $j$ of a newly activated node $i$ is likely to also be a neighbour of the nodes that influenced $i$ to adopt, and so $j$ has an enhanced chance of itself becoming an adopter.
If a suitable neighbourhood is present under these circumstances, one uptake will then  trigger another in an adjacent position due to this overlap of neighbourhoods (see Fig.~\ref{fig:clique}).
If clustering is sufficient that there is overlap between neighbourhoods for a sizeable portion of the whole network, then it can  take only a single critical neighbourhood to induce a successful uptake for the system. 
The 50\% chance of success seen in Figure \ref{fig:plotsN} (a) in the $Y^*=3$ sector then represents the chance of randomly seeding a group of nodes in such a way as to trigger this process.
The probability given by (\ref{eqn:PZ1}), of at least one node in a random system having a critical neighbourhood, yields $P(Z \geq 1)\approx 0.67$ for $k=6$, $Y^*=3$ and $N=500$.
Equation (\ref{eqn:PZ1}) is strictly valid only for independent neighbourhoods, and hence gives too high a value in this case, where neighbourhoods overlap, but a quantitatively accurate expression would require a far more intensive calculation of the different combinations of the initial seed for the whole system. 
However, the current method is accurate enough to give qualitatively correct predictions.

The propagation via overlapping neighbourhoods is a percolation phenomenon, similar to that seen in other studies \cite{newman2003properties}, but percolating on \emph{hyper-edges} connecting \emph{hyper-nodes}, rather than via single links.
Here, a hyper-node is a complete sub-graph of $Y^*+1$ nodes, or a \emph{$(Y^*+1)$-clique}.
Two hyper-nodes are connected if the two ($Y^*+1$)-cliques have $Y^*$ nodes in common.
The implication is that if a hyper-node becomes active, then every hyper-node to which it is connected will also become active in the next time-step.  
This is shown in Figure \ref{fig:clique}, for propagation via 4-cliques where $Y^*=3$, with a single uptake triggering others via their overlapping neighbourhoods.

\begin{figure}[h!]
\centering
(a)\includegraphics[width=0.5\linewidth]{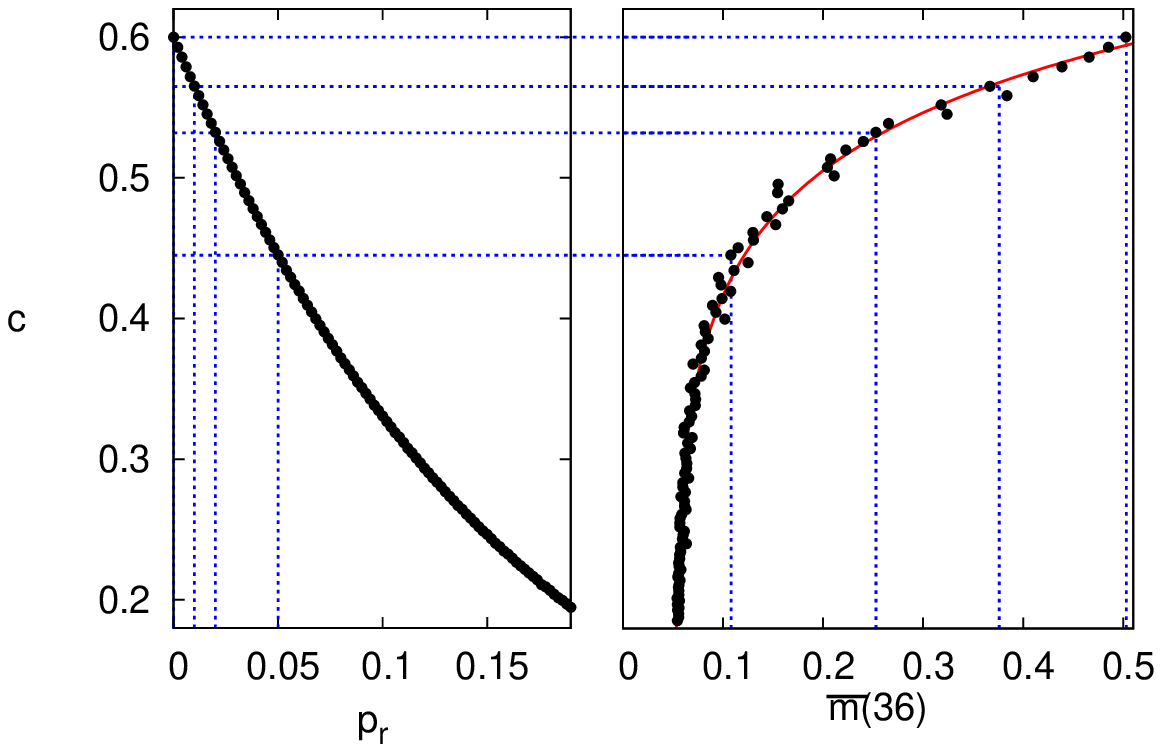}
(b)\includegraphics[width=0.4\linewidth]{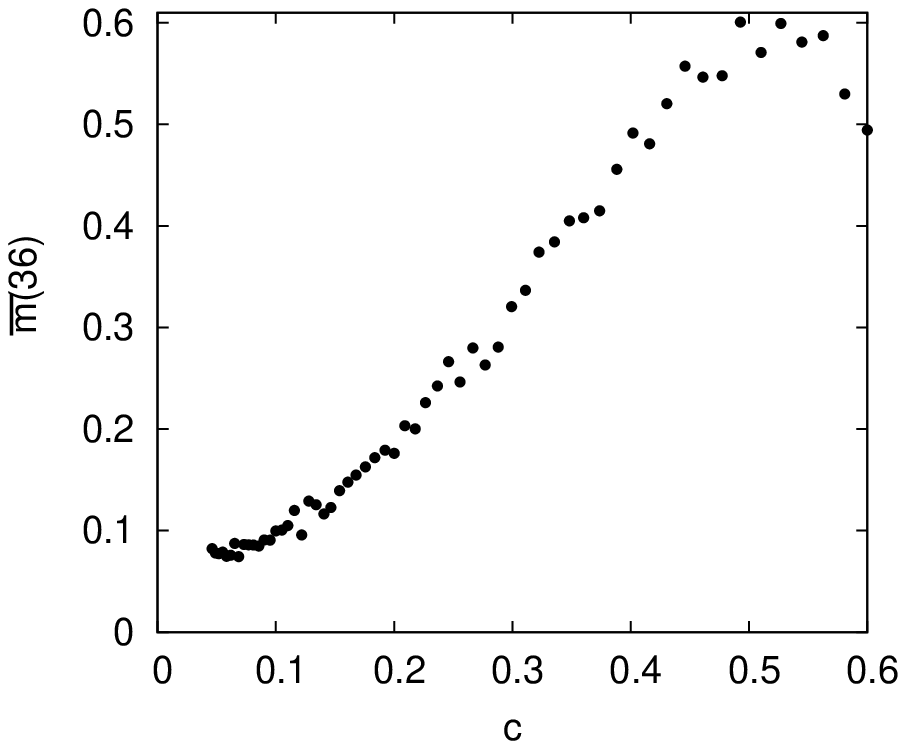}
\caption{Uptake on Watts--Strogatz rewired 6-neighbour rings, after 36 time-steps, against both rewiring probability $p_r$ and clustering $c$. (a) Runs starting from 1000  random initial conditions with $m_0=0.05$, $\alpha=0.1, \beta=0.45, \gamma=0.45, p=0.5, \theta=0.25$, randomising the network on each realisation.  Edges are rewired in pairs to preserve the node degree $k=6$. Dashed blue lines show the data in Fig.~\ref{fig:plotsN}, with $p_r=0,0.01,0.02,0.05$ and $c=0.6,0.57,0.53,0.45$, respectively. The fitted red curve is the function $\sigma \exp(\varepsilon c)$, with $\sigma=2.9\times10^{-4}$, $\varepsilon=12.39$. (b) For comparison, the same system is used but with edges rewired individually between $0<p_r<0.5$, showing the effect of allowing distribution of the node degree with $p_r$.  A peak in the uptake appears at $c\approx0.5$.}\label{fig:clust}
\end{figure}

To demonstrate the transition between cluster-enabled propagation and stagnation, the ensemble uptake within the $Y^*=3$ sector of Figure \ref{fig:plotsN} (after 36 time-steps) is shown in Figure \ref{fig:clust}, plotted as a function of both $p_r$ and $c$.
The regular (un-rewired) network has a clustering coefficient $c=0.6$, which decreases smoothly with increasing rewiring probability.
If the node degree is held constant by rewiring edges in pairs (Fig.~\ref{fig:clust} (a)) then the average uptake after 36 time-steps fits well with the function $m(36)=\sigma \exp(\varepsilon c)$, with fitting parameters $\sigma=2.9\times10^{-4}$ and $\varepsilon=12.39$.  This is not the case if edges are rewired one at a time (Fig.~\ref{fig:clust} (b)), where the node degree becomes distributed.
In this case the uptake peaks at around 5\% rewiring ($c\approx0.5$), similar to other models showing a peak in the level of adoption at certain node degrees \cite{alkemade2005strategies}.

\begin{figure}
\centering
(a)\includegraphics[width=0.45\linewidth]{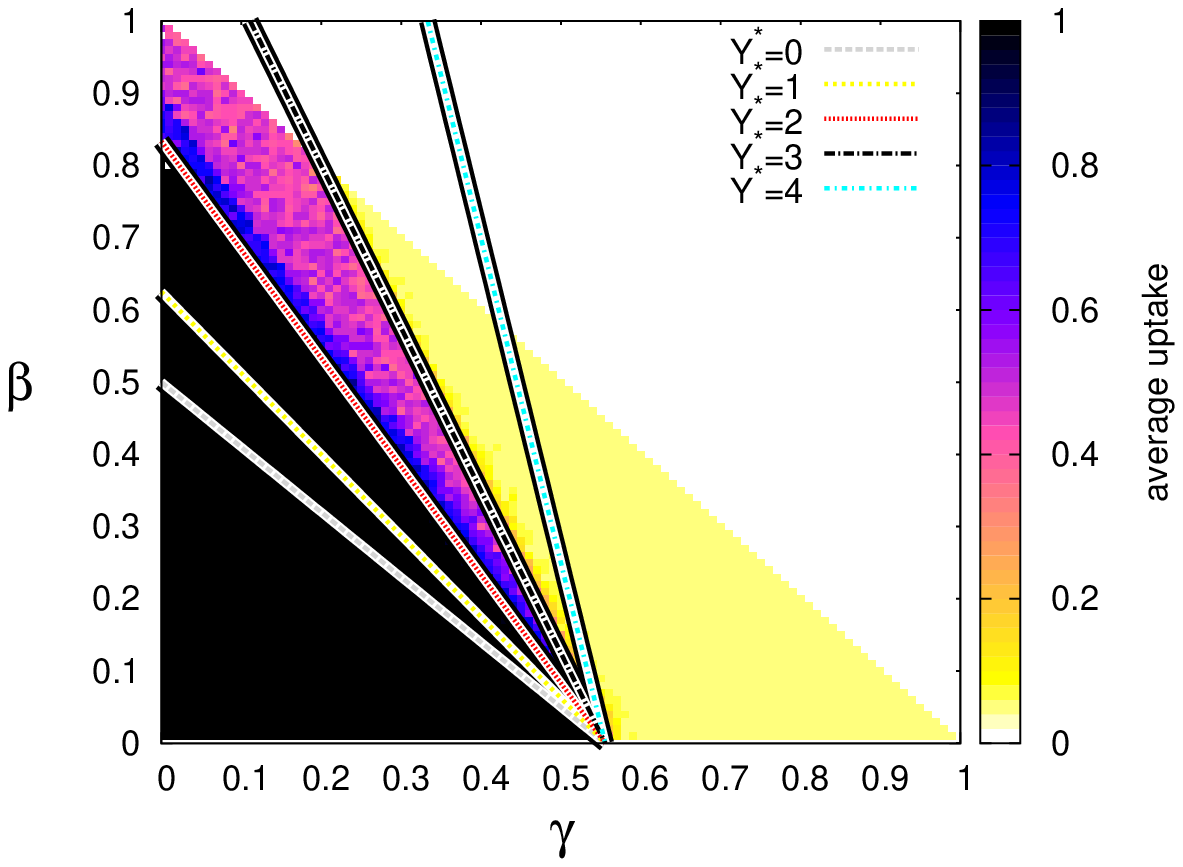}
(b)\includegraphics[width=0.45\linewidth]{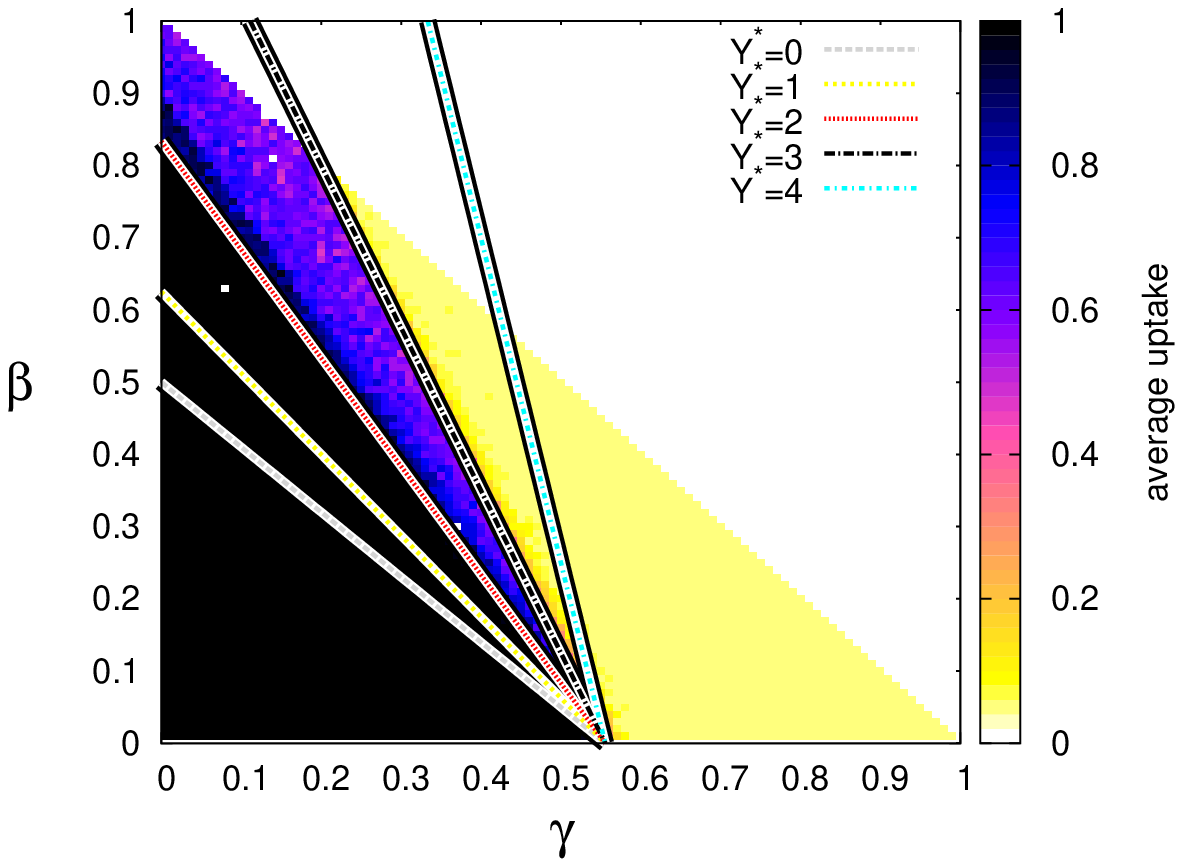}
(c)\includegraphics[width=0.45\linewidth]{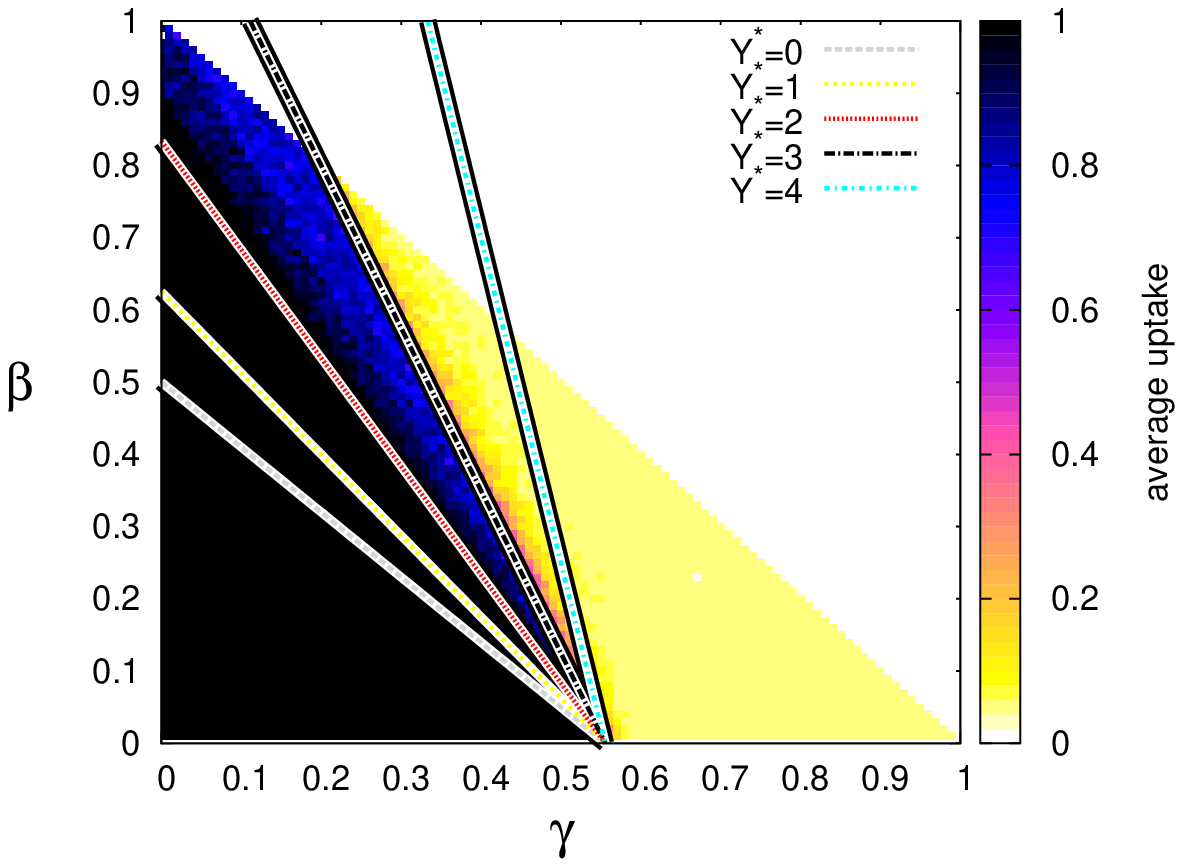}
(d)\includegraphics[width=0.45\linewidth]{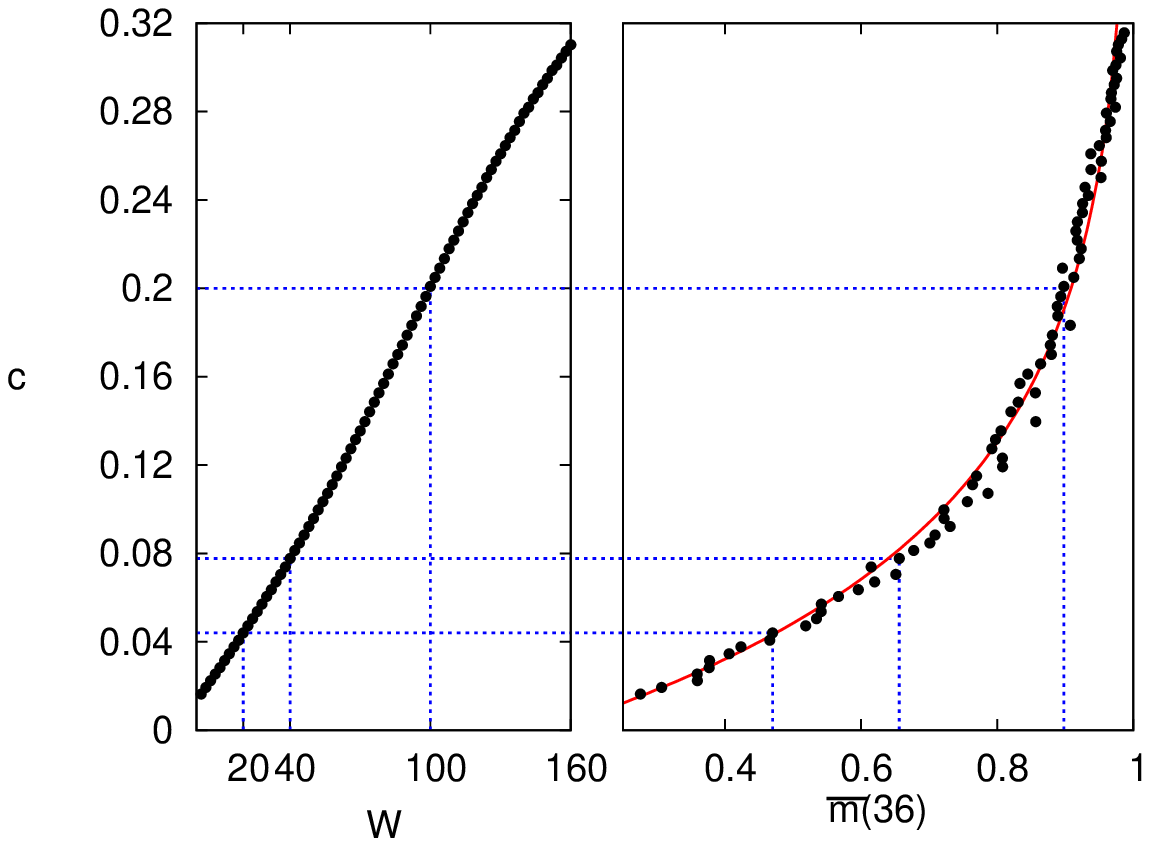}
\caption{Same as Figs.~\ref{fig:plotsN} and \ref{fig:clust} but for a community-based, Newman random clustered, network model. The network size $N=500$, so for each number of groups $W$ there are $M$ members, on average, resulting in a measured transitivity $c$: (a) $W=20$, $M=25$, $c=0.044$ (b) $W=40$, $M=12.5$, $c=0.078$ (c) $W=100$, $M=5$, $c=0.2$ (d) Clustering as a function of W versus uptake for $\alpha=0.15,\beta=0.8,\gamma=0.05$ within the $Y^*=3$ sector. The fitted red line is the empirical function $m(36)=1-\sigma \exp(-\varepsilon c)$, with $\sigma=0.86$, $\varepsilon=11.2$.}\label{fig:newclust}
\end{figure}

A further example is shown in Figure \ref{fig:newclust} for a community based, random clustered network \cite{newman2003properties}, where the clustering can be varied by associating $N$ individual nodes via $G$ groups of varying sizes $M$.
The number of communities $W$ and number of connections $L$ per community (here $L=5$) determines the size and density of the communities and hence the amount of clustering.
In Figure \ref{fig:newclust} (d) the dependence of $c$ on $W$ is shown alongside the mean uptake after 36 time-steps, averaged over 1000 realisations for each $W$.
Similar results can be seen as in the previous case, with increased clustering resulting in an increased likelihood of successful uptake, despite the low number of influential neighbourhoods.
However, these expected values are only valid when averaging over different network realisations as well as initial adopters.
For many networks it was found that uptake was almost never achieved for most choices of initial adopters, which we believe to be related to the connectedness of cliques of size $Y^*+1$.
However, an interesting observation is the strong functional dependence of the expected success on the \emph{transitivity}, which fits to about 1\% accuracy to the function:
\begin{equation}
m(36)=1-\sigma \exp(-\varepsilon c),\label{eqn:fit}
\end{equation}
with fitting parameters $\sigma=0.86$, $\varepsilon=11.2$.
This relationship was also found for different values of $(\alpha,\beta,\gamma)$ within the $Y^*=3$ sector, with different coefficients in (\ref{eqn:fit}).
This suggests that it is  possible to obtain an analytical expression for these systems, and should motivate further research.

The phenomenon of connectedness via more highly connected clusters than single edges is known as $k$-clique percolation, and results exist deriving this result for random (ER) networks \cite{derenyi2005clique,palla2007critical}.
However,  to understand the results in Figure \ref{fig:newclust} analytically would require results for the expected connected $(Y^*+1)$-clique percolation component size distribution as a function of the clustering coefficient.
Some progress has been made in this direction, with numerical results existing for cascades of various models, such as Watts' threshold model \cite{watts2002simple} on clustered random networks like the Newman model \cite{hackett2011cascades}. 
Additionally analytical results have also been derived for expected connected component sizes for single bond percolation on clustered networks \cite{gleeson2009bond}.
However, obtaining the required results for $k$-clique percolation components remains an open and difficult problem.

The results in this section imply that, in cases where several network neighbours are required to influence adoption, greater levels of system-wide adoption can be achieved in more highly clustered networks, in agreement with earlier results on similar systems \cite{choi2010role}.
Therefore, innovations in systems with more communication between overlapping groups of individuals will be more likely to succeed than otherwise.  

\section{Conclusions}\label{sec:conc}

Numerical simulations have been carried out, computing the average uptake of an innovation by householders connected by a network of peer to peer influences. 
This was carried out over many realisations, at a range of model parameter values, on various network topologies and initial conditions.
It has been found that the expected level of adoption depends strongly on the topology of the network and the $Y^*$ sector that the $(\beta,\gamma)$ parameters are in. 
The node degree and the network's transitivity, related to the correlation between the neighbourhoods of connected individuals, were found to be of particular importance.

An analytical and probabilistic approach has been developed to account for the observed behaviour.
The analysis explains the results of the numerical calculations well in the case of randomly connected networks, and provides useful insight in the case of networks with clustering.

The model provides a basis to simulate the behaviour seen in the real world.
For example, it could form the basis for computational investigations comparing different interventions by a government to try to improve the level of uptake of innovations relating to energy efficiency.
These could be represented in the model by varying the initial seeding for different roll-out strategies, or network structure for different word-of-mouth marketing campaigns.
By identifying intervention strategies that have a high probability of achieving successful uptake of the technology, the model could allow government to make informed decisions of whether to go ahead with interventions and if so, how.
This will be the subject of a future paper.

Future studies can modify the community-based network models to provide more flexibility and realistic features in a number of ways.
Some positional information can be ascribed to the nodes and some or all of the groups, preferentially attaching individuals to communities that are geographically closer, in a similar way to Hamill and Gilbert's \emph{Social Circles} model \cite{hamill2009social}.
In cases where none (or only one) of the groups have geographic information they again reduce to the Newman scheme.
These groups could be classified into different categories, for example some could have a fixed membership number to allow variation of the clustering between groups and represent different types of group interaction (e.g., social or workplace groups).
For other types there could be no geographical element, introducing long-range connections into the network, as may be the case for workplaces.

The analysis of the model made use of the assumption of a near-constant node degree, and so would not carry over simply to scale-free networks.
In this work we argued that communication on innovation adoption is restricted to a select few contacts, therefore excluding wide degree distributions.
An alternative would be to assume that individuals are influenced by a \emph{number} rather than \emph{fraction} of their network neighbours, which is independent of $\bar k$.

The dynamical adoption model could  be made to account for the various archetypes discussed in the social science literature by assigning individuals to different groups.
Individuals in different categories would be assigned different $\alpha_i$, $\beta_i$ and $\gamma_i$ values, depending on their personality, for example with technology enthusiasts having high $\alpha$ values, social followers having high $\beta$, and cautious individuals having a high $\gamma$.
The model could also be given a stochastic element by allowing a distribution of time-step values, and introducing a probabilistic, rather than deterministic, uptake once the utility exceeds its threshold.

The methods discussed in this paper can also be applied to multi-parameter models of different systems with similar features, and provide a framework for the investigation of many real-world systems.

\section*{Acknowledgements}
  
We thank Mason Porter for some  very useful discussions which led us on the path to deeper understanding of this problem from the probability perspective.	
	
\bibliography{ref}
\bibliographystyle{siam.bst}

\end{document}